\documentclass[10pt,aps,prd,twocolumn,nofootinbib,preprintnumbers,superscriptaddress,showkeys,floatfix,amsmath,amssymb]{revtex4-2}
\usepackage{graphicx}
\usepackage{amsmath,amssymb}
\usepackage{amsfonts}
\usepackage{xspace} 
\usepackage[usenames]{color}
\usepackage{bm}
\usepackage{mathrsfs}
\usepackage[colorlinks=true]{hyperref}
\usepackage[all]{hypcap} 
\usepackage[utf8]{inputenc} 
\usepackage{slashed}
\usepackage{rotating}
\usepackage{nicematrix}
\usepackage{makecell}
\usepackage[bbgreekl]{mathbbol}
\usepackage{relsize}
\usepackage[normalem]{ulem}
\DeclareSymbolFontAlphabet{\mathbb}{AMSb}
\DeclareSymbolFontAlphabet{\mathbbl}{bbold}

\usepackage{tikz,xcolor,hyperref}
\definecolor{lime}{HTML}{A6CE39}
\DeclareRobustCommand{\orcidicon}{
	\begin{tikzpicture}
	\draw[lime, fill=lime] (0,0) 
	circle [radius=0.16] 
	node[white] {{\fontfamily{qag}\selectfont \tiny ID}};
	\draw[white, fill=white] (-0.0625,0.095) 
	circle [radius=0.007];
	\end{tikzpicture}
	\hspace{-2mm}
}

\foreach \x in {A, ..., Z}{%
	\expandafter\xdef\csname orcid\x\endcsname{\noexpand\href{https://orcid.org/\csname orcidauthor\x\endcsname}{\noexpand\orcidicon}}
}

\newcommand{\orcid}[1]{\href{https://orcid.org/#1}{\textcolor[HTML]{A6CE39}{\aiOrcid}}}

\begin{document}

\title{Bayesian inference favors quark matter in neutron star interiors}



\author{Alexander Ayriyan\orcidA{}}
\email{alexander.ayriyan@uwr.edu.pl}
\affiliation{
Institute of Theoretical Physics, University of Wroclaw, Max Born Pl. 9, 50-204, Wroclaw, Poland}
\affiliation{
A. Alikhanyan National Science Lab, 
Alikhanyan Brothers street 2, 0036 Yerevan, Armenia}

\author{Oleksii Ivanytskyi\orcidB{}}
\email{oleksii.ivanytskyi@uwr.edu.pl}
\affiliation{Incubator of Scientific Excellence---Centre for Simulations of Superdense Fluids, University of Wrocław, 50-204, Wroclaw, Poland
}

\author{David Blaschke\orcidC{}}
\email{david.blaschke@uwr.edu.pl}
\affiliation{
Institute of Theoretical Physics, University of Wroclaw, Max Born Pl. 9, 50-204, Wroclaw, Poland}
\affiliation{Helmholtz-Zentrum Dresden-Rossendorf (HZDR), Bautzner Landstrasse 400, 01328 Dresden, Germany}
\affiliation{Center for Advanced Systems Understanding (CASUS), Untermarkt 20, 02826 G\"orlitz, Germany}

\date{\today}
\begin{abstract}
We perform a physics-informed Bayesian analyses of the equation of state of hybrid neutron stars that incorporates color-flavor-locked quark matter modeled by a three-flavor non-local Nambu-Jona-Lasinio framework with vector repulsion and diquark pairing. Contrary to the model-agnostic Bayesian analyses our scheme allows for distinguishing between the scenarios of neutron stars with quark cores and without them. The used quark model realizes asymptotic conformality at high densities in accordance with perturbative QCD. The hadronic sector is described by the density-dependent relativistic functional DD2Y-T, which satisfies chiral effective field theory constraints and includes hyperonic degrees of freedom. We construct a large set of candidate hybrid EOSs by varying the vector and diquark couplings and apply a Maxwell construction for the quark–hadron phase transition. Observational constraints from recent NICER pulsar mass–radius measurements and tidal deformability from GW170817 are incorporated into the likelihood. 
Depending on whether the observational data from the black widow pulsar PSR J0952-0607 and the HESS J1731-347 object are included to the analysis or not,
the posterior distribution favors vector and diquark couplings around $(\eta_V,\eta_D)\simeq (0.82,0.40)$ or $(\eta_V,\eta_D)\simeq (0.64,0.36)$, respectively. 
This corresponds to equations of state that support two-solar-mass neutron stars with superconformal speed of sound and relatively low onset densities for deconfinement. Our findings indicate that the most probable hybrid EOSs are statistically preferred over the purely hadronic baseline.
The corresponding probabilities of agreeing with the observational data differ by one or two orders of magnitude depending on the data set used. This suggests that quark cores may exist in all observed neutron stars.
\end{abstract}
\maketitle

\section{Introduction}
\label{intro}

Neutron stars (NSs) can be considered as astrophysical laboratories for the investigation of strongly interacting matter under conditions of high baryon densities and low temperatures, which are not accessible in terrestrial experiments, e.g., with ultrarelativistic heavy-ion collisions. 
One of the key questions still debated controversially is whether the inner core of NSs may consist of deconfined quark matter. 
In other words, is the critical mass $M_{\rm onset}$, where quark deconfinement sets in, smaller than the maximum mass of the sequence of stable NSs for a given equation of state?
A positive answer to this question would imply that the critical density for the onset of deconfinement under the NS conditions is low enough to be attained. 

The most striking and unambiguous signal for a strong phase transition in the NS interior would be the observation of mass twin NSs, see \cite{Blaschke:2013ana,Benic:2014jia} and references therein. 
The detection of such a pair of stars with almost identical masses but significantly different radii requires capabilities to measure mass and radius of pulsars simultaneously with sufficient accuracy. The best instrument of such a kind presently operating is the NICER X-ray observatory mounted on the International Space Station. 
While no twins have yet been detected, the information on masses and radii of millisecond pulsars in a wide range of masses from $\sim 1.4\rm M_\odot$ to $\sim 2.0 \rm M_\odot$ together with the compactness constraints from the tidal deformability measurement on the binary NS merger GW170817 by the LIGO-Virgo Collaboration \cite{LIGOScientific:2017vwq} provides fragments of the mass-radius relation for NSs, which by inversion of the Tolman-Oppenheimer-Volkoff (TOV) equations, is equivalent to measuring the NS equation of state (EOS). 
In order to infer this EOS from the still error-prone data for the mass and radius constraints, Bayesian analysis methods are the appropriate tool. 
One can use them in the agnostic manner, where the most likely EOS is selected as a posterior among a huge number of algorithmically sampled curves in the pressure vs. energy density plane (see, e.g., \cite{Brandes:2022nxa,Brandes:2023hma}) or one can apply a physics-informed scheme, where a generic form of the EOS with physically meaningful parameters is suggested and just these parameters are varied within reasonable limits (meta-modeling), see \cite{Ayriyan:2024zfw,Li:2025tku,Grundler:2025mcz} for examples that implement a hadron-to-quark matter phase transition.

In the present work, we apply the physics-informed Bayesian analysis scheme developed in Ref. \cite{Ayriyan:2024zfw} to investigate the question whether hybrid EOS is preferable over a baseline hadronic EOS under present-day mass-radius constraints for pulsars. 
The main advantage of this scheme compared to the widely used model-agnostic Bayesian analyses \cite{Annala:2019puf,Altiparmak:2022bke,Marczenko:2022jhl,Brandes:2022nxa,Brandes:2023hma} is the possibility to distinguish between the scenarios of NSs with quark cores and without them. These scenarios can be characterized by the posterior probabilities of agreeing with the observational data.
Comparing these probabilities allows to quantitatively answer the question whether the NS cores may contain deconfined quark matter.
Below we analyze these probabilities  for the first time and show that according to the existing observational data on the mass-radius relation of NSs and their tidal deformability the existence of quark cores in NSs is from one to two orders of magnitude more probable than the absence of them.

Compared to the previous physics-informed Bayesian analysis performed in Ref. \cite{Ayriyan:2024zfw}, we apply the following qualitative improvements of the scheme towards a most realistic description of dense NS matter.
For the quark matter EOS we employ a recently developed three-flavor nonlocal chiral quark model of the Nambu--Jona-Lasinio (NJL) type with color-flavor-locked (CFL) color superconductivity and a nonlocal vector repulsion of the mean-field type, which assures that at asymptotically large densities the conformal limit is reached \cite{Ivanytskyi:2024zip}. 
As the hadronic baseline EOS we use the relativistic density functional approach DD2npY-T recently depeloped by Stefan Typel \cite{Shahrbaf:2022upc}.
For the Bayesian analysis, we include the new NICER mass-radius constraint  for the $1.4\rm M_\odot$ pulsar PSR J0614--3329 by Mauviard et al. \cite{Mauviard:2025dmd}.

The paper is organized as follows. 
In Section~\ref{sec2}, we introduce the nonlocal chiral quark model for the CFL phase of quark matter with special emphasis on analyzing its high density asymptotic, consistent with the conformal limit of QCD.
In Section~\ref{sec4}, we describe the set of EOS used in the analysis, while the Bayesian framework is outlined in Section~\ref{sec5}. In Section~\ref{sec6} we present our results and their discussion, Section~\ref{concl} gives the Conclusions. 

\section{Asymptotically conformal CFL quark matter}
\label{sec2}

In this work we consider the scenario of hybrid NSs with hadronic outer layers and quark inner core. 
The EOS of hadron matter is modeled within the DD2Y-T model supplemented with a crust EOS \cite{Shahrbaf:2022upc}.
The degrees of freedom of this model include nucleons, hyperons and electrons, which provide electric neutrality of the hadronic EOS.
These components exist in $\beta$-equilibrium.
The model respects the low density constraint of the chiral effective field theory ($\chi$EFT) \cite{Kruger:2013kua} and reproduces the most important properties of the nuclear matter ground state.

The EOS of quark cores is modeled within the recently proposed three-flavor nonlocal NJL model with vector repulsion and CFL color-superconductivity \cite{Ivanytskyi:2024zip} treated within the separable approximation \cite{Blaschke:1994px,GomezDumm:2001fz}.
The model is formalized by the Lagrangian 
\begin{eqnarray}
\mathcal{L}&=&\overline{q}(i\slashed\partial-m+\mu\gamma_0)q\nonumber\\
\label{I}
&+&G_S\hspace*{-.1cm}\sum_{a=\overline{0,8}}\hspace*{-.1cm}s_as_a-G_Vj_\mu j^\mu
+3G_D\hspace*{-.3cm}\sum_{a,b=2,5,7}\hspace*{-.3cm}d_{ab}^+d_{ab}^{ },
\end{eqnarray}
where $q=(u,d,s)^{T}$ is three-flavor quark field, $m$ denotes the current mass being the same for all quark flavors and $G_S$, $G_V$ and $G_D$ stand for the coupling constants controlling the strength of interaction in the scalar, vector and diquark pairing channels.
These channels have four-quark form and include nonlocal quark currents
\begin{eqnarray}
\label{II}
s_a(x)&=&\int dz~g(z)\overline q\left(x+\frac{z}{2}\right)\tau_aq\left(x-\frac{z}{2}\right),\\
\label{III}
j_\mu(x)&=&\int dz~g(z)\overline q\left(x+\frac{z}{2}\right)\gamma_\mu q\left(x-\frac{z}{2}\right),\\
\label{IV}
d_{ab}^{ }(x)&=&\int dz~g(z)\overline q\left(x+\frac{z}{2}\right)i\gamma_5\tau_a\lambda_bq^c\left(x-\frac{z}{2}\right),
\end{eqnarray}
where $\tau_0=\sqrt{2/3}$ is introduced along with the flavor Gell-Mann matrices $\tau_{a\ge1}$ for the sake of unifying the notations, $\lambda_b$ are color Gell-Mann matrices, $q^c=i\gamma_2\gamma_0\overline{q}^T$ represents charge conjugated quark field and $g(z)$ is space-time dependent formfactor.
In Ref. \cite{Ivanytskyi:2024zip} this formfactor was chosen in the instantaneous Gaussian form so that its time dependence reduces to $\propto\delta(x_0)$, while its three-momentum dependent Fourier transform reads
\begin{eqnarray}
    \label{eq:formfactor}
    g_{\bf k}=\exp(-{\bf k}^2/\Lambda^2),
\end{eqnarray}
where constant $\Lambda$ defines the scale separating low-momentum non-perturbative and high momentum perturbative modes.

It is important to stress that chirally symmetric form of the Lagrangian (\ref{I}) also includes the pseudoscalar current densities.
It can be obtained by replacing $\tau_a$ in Eq. (\ref{II}) by $i\gamma_5\tau_a$.
However, the corresponding currents vanish under the mean-field approximation considered in Ref. \cite{Ivanytskyi:2024zip}.
As a result, the pseudoscalar channel does not play a role in using the present model for constructing EOS of quark matter.
Therefore, it is omitted for shortening the notations.

The described model suggests a simplified picture of three-flavor quark matter when light and strange quarks are degenerated in mass, which corresponds to the so-called CFLL quark matter \cite{Blaschke:2023}.
This simplification is related to the possibility of finding analytically the single-quark energies as eigen-values of the $72\times72$ Nambu-Gorkov propagator (see Ref. \cite{Buballa:2003qv} for details).
At the same time, high physical masses of strange quarks can lead to their suppression relative to light quarks and consequent destruction of the CFL color-superconductivity, which requires equal amounts of all quark flavors. 
In Ref. \cite{Ivanytskyi:2024zip} stability of quark matter with respect to {\it "unlocking"} of strange quarks from the CFL phase was analyzed based on a simple kinematic criterion \cite{Schafer:1999pb,Alford:1999pa,Abuki:2003ut} and proven to be the case for the physically relevant values of the model parameters.
This allows us to conclude that the proposed model of the CFLL color-superconductivity provides a reasonably accurate description of cold three-flavor quark matter.
It is also important to stress, that equality of masses of light and strange quarks provides electric neutrality and $\beta$-equilibrium of the CFLL quark matter even in the absence of electrons.

The thermodynamic potential of the present model can be found by performing the sequence of the steps described in Ref. \cite{Ivanytskyi:2024zip}.
For the convenience of the readers we briefly repeat them in Appendix \ref{appA}.
The first of them corresponds to bosonizing the Lagrangian (\ref{I}) by applying the Hubbard-Stratonovich transformation of the partition function. 
This introduces to the consideration auxiliary bosonic fields $\sigma_a$, $\omega_\mu$ and $\Delta_{ab}$, which are conjugated to the scalar, vector and diquark interaction channels, respectively.
At the second step the mean field approximation is applied, i.e., the bosonic fields are replaced by their expectation values found from the corresponding averaged Euler-Lagrange equations.
Only the zeroth scalar field does not vanish under the averaging, i.e., $\langle\sigma_a\rangle=\delta_{a0}\sigma$.
It is related to the spontaneous breaking and dynamical restoration of chiral symmetry since $\sigma$ enters the effective quark mass $M_{\bf k}=m+\sigma g_{\bf k}$\footnote{\label{note1}
As explained in the erratum \cite{PhysRevD.111.079904}, Ref. \cite{Ivanytskyi:2024zip} includes a rescaling of the scalar coupling $G_S$ by the factor $2/3$, which is not mentioned but, nevertheless, does not affect the model results.
Without this rescaling the effective quark mass reads $M_{\bf k}=m+\sqrt{2/3}\sigma g_{\bf k}$.
In this work we follow the notations of Ref. \cite{Ivanytskyi:2024zip}.} as the corresponding mass gap.
Furthermore, only the zeroth component of the vector field, which quantifies the effects of vector repulsion, survives so that $\langle\omega_\mu\rangle=g_{\mu0}\omega$.
Similarly, only the diagonal diquark fields attain the same nonvanishing expectation value $\Delta\equiv\langle\Delta_{11}\rangle+\langle\Delta_{22}\rangle+\langle\Delta_{33}\rangle$.
This value plays the role of the diquark pairing gap and serves as the order parameter of the phase transition between normal quark matter and its CFL phase.
With this the mean-field thermodynamic potential of cold CFLL quark matter reads
\begin{eqnarray}
\Omega&=&-\sum_{j,a}d_j\int\frac{d\bf k}{(2\pi)^3}\left[\frac{1}{2}-
\theta\left(-\epsilon_{j\bf k}^a\right)\right]\epsilon^a_{j\bf k}
\nonumber\\
\label{eq:Omega}
&+&\frac{\sigma^2}{4G_S}-\frac{\omega^2}{4G_V}+\frac{\Delta^2}{4G_D}.
\end{eqnarray}
It includes the single particle energies of quarks ($a=+$) and antiquarks ($a=-$) shifted by the quark chemical potential
\begin{eqnarray}
\label{eq:dispersion}
\epsilon^a_{j{\bf k}}={\rm sign}\left(\epsilon_{\bf k}^a\right)
\sqrt{{\epsilon_{\bf k}^a}^2+\Delta_{j\bf k}^2}.
\end{eqnarray}
Here $\epsilon_{\bf k}^\pm=\sqrt{{\bf k}^2+M_{\bf k}^2}\mp\mu\mp\omega g_{\bf k}$ and $\Delta_{j\bf k}=\zeta_j \Delta g_{\bf k}$ with $\zeta_{\rm sing}=2$ and $\zeta_{\rm oct}=1$.
The subscript index $j$ distinguishes between the singlet and octet states \cite{Buballa:2003qv,Blaschke:2023,Ivanytskyi:2024zip}.
The corresponding degeneracy factors are $d_{\rm sing}=d/9$ and $d_{\rm oct}=8d/9$ with $d=2N_cN_f$ being the spin-color-flavor degeneracy of quarks.

Equivalently to averaging the Euler-Lagrange equations, the expectation values of the scalar, vector and diquark fields can be found by minimizing the thermodynamic potential (\ref{eq:Omega}).
Having these expectation values found, the regularized thermodynamic potential $\Omega^{\rm reg}$ is obtained by subtracting from $\Omega$ the constant diverging thermodynamic potential of free quarks in the vacuum, i.e.,
\begin{eqnarray}
\label{eq:Omegareg}
{\Omega^{}}^{\rm reg}&=&\Omega+d\int\frac{d\bf k}{(2\pi)^3}\sqrt{{\bf k}^2+m^2}.
\end{eqnarray}

The effective bag pressure $B_{\rm eff}$ is a well known phenomenological parameter of quark matter.
Within the chiral quark models of the NJL type
it can be obtained as the negative of the vacuum value of the regularized thermodynamic potential (see e.g. Ref. \cite{Buballa:2003qv}).
For the present model this gives
\begin{eqnarray}
    \label{VIII}
    B_{\rm eff}&=&d\int\frac{d{\bf k}}{(2\pi)^3}\left[\sqrt{M_{0,{\bf k}}^2+{\bf k}^2}-
    \sqrt{m^2+{\bf k}^2}\right]\nonumber\\
    &-&\frac{(M_{0,{\bf k}=0}-m)^2}{4G_S}.
\end{eqnarray}
Hereafter the subscript index ``$0$'' denotes the quantities defined in the vacuum.
Eq. (\ref{VIII}) allows us to connect the vacuum value of the effective quark mass at vanishing momentum to the effective bag pressure.
Higher values of $M_{0,{\bf k}=0}$ correspond to higher values of $B_{\rm eff}$, which is a reflection of the fact that dynamical generation of the quark mass gap and confining bag pressure are related through the phenomenon of spontaneous chiral symmetry breaking.
 
The pressure, baryon and energy densities of cold CFLL quark matter can be obtained from the regularized thermodynamic potential as $p=-\Omega^{\rm reg}-B_{\rm eff}$, $n_B=\partial p/\partial\mu_B$ and $\varepsilon=\mu_Bn_B-p$, respectively.
This definition provides the vanishing vacuum pressure, while $\mu_B$ is baryonic chemical potential.
Below we also consider the squared speed of sound $c_S^2=\partial p/\partial\varepsilon$ and the dimensionless interaction measure $\delta=1/3-p/\varepsilon$, which are used to quantify the stiffness of the NS matter and its deviation from the conformal behavior.

The typical strategy of fixing the parameters of chiral quark models of the NJL type is to fit masses and decay constants of the pseudoscalar mesons \cite{Vogl:1991qt,Klevansky:1992qe,Hatsuda:1994pi,Buballa:2003qv,Ratti:2005jh,Ratti:2006wg,CamaraPereira:2016chj,Ivanytskyi:2022oxv}.
However, this strategy can not be directly applied in a three-flavor model with mass degeneration of all the flavors.
Therefore, in Ref. \cite{Ivanytskyi:2024zip} an alternative scheme, aiming at finding reasonable values of the model parameters rather than at their ultimate determination, was suggested.
The current quark mass was fixed as the mean value of the current masses of $u$- and $d$-quarks from the Review of Particle Physics \cite{ParticleDataGroup:2022pth}, i.e. $m=(m_u+m_d)/2=3.5$ MeV.
The momentum scale $\Lambda$ and scalar coupling $G_S$ were determined according to the vacuum values of the effective quark mass at vanishing momentum $M_{0,{\bf k}=0}$ and chiral condensate per flavor $\langle\overline{f}f\rangle_0$. 
Following Ref. \cite{Ivanytskyi:2024zip}, we fix $\langle\overline{f}f\rangle_0=-(250~\rm MeV)^3$, which is a compromise between the values of the QCD sum rules at the renormalization scale of 1 GeV, i.e., $-(229\pm33~\rm MeV)^3$ \cite{Dosch:1997wb} and $-(242\pm15~\rm MeV)^3$ \cite{Jamin:2002ev}, and the prediction made according to the Gell-Mann-Oakes-Renner relation \cite{Gell-Mann:1968hlm} with the pion mass and decay constant from the Review of Particle Physics \cite{ParticleDataGroup:2022pth} and the current quark mass mentioned above, i.e., $-(280\pm12~\rm MeV)^3$.
Given a gauge dependence of the momentum-dependent quark mass extracted from their propagator and strong impact of the current quark mass, we use the lattice QCD result $M_{0,{\bf k}=0}=300-500$ MeV \cite{Skullerud:2000un,Skullerud:2001aw,Parappilly:2005ei,Burgio:2012ph} as a guidance and consider the mean value of the vacuum effective quark mass at vanishing momentum, i.e.,  $M_{0,{\bf k}=0}=400$ MeV. 
The corresponding values of the model parameters along with the effective bag pressure are shown in Table \ref{table1}.
The vector and diquark couplings are parameterized by the dimensionless parameters $\eta_V=G_V/G_S$ and $\eta_D=G_D/G_S$, which below are referred to as couplings. 
Having the parameters given in Table \ref{table1} fixed, the EOS of quark matter is fully determined by $\eta_V$ and $\eta_D$.

\begin{table}[t]
\begin{tabular}{|c|c|c|c|c|c|}
\hline    
      $m$    &  $\Lambda$  &     $G_S$        &  $M_{0,{\bf k}=0}$ & $|\langle\overline{f}f\rangle_0|^{1/3}$ & $B_{\rm eff}$     \\
 $\rm [MeV]$ & $\rm [MeV]$ & $\rm [GeV^{-2}]$ &    $\rm [MeV]$     &                 $\rm [MeV]$             &  $[\rm MeV/fm^3]$ \\ \hline
      3.5    &    564      &       9.62       &          400       &                    250                  &       81.3        \\ \hline
\end{tabular}
\caption{Values of the model parameters used in this work and the resulting effective mass at zero momentum, chiral condensate per flavor in the vacuum and the effective bag pressure.} 
\label{table1}
\end{table}

At high densities typical quark momenta significantly exceed $\Lambda$.
In this case interaction among quarks is strongly suppressed by the formfactor (\ref{eq:formfactor}).
The quark masses also can be neglected at high densities.
As a result, dense CFLL quark matter behaves as a gas of weakly interacting massless particles.
Therefore, the CFLL quark matter exhibits asymptotically conformal behavior with the speed of sound and dimensionless interaction measure, behaving as $c_S^2=1/3-3m^2/\mu_B^2$ and $\delta=3m^2/\mu_B^2$ \cite{Ivanytskyi:2024zip}.
Remarkably, the obtained asymptotic expressions coincide with the ones corresponding to cold QCD matter at zero baryon density but finite isospin asymmetries \cite{Ivanytskyi:2025cnn} if $\mu_B/3$ is treated as the quark chemical potential. 
At the same time, if the vector repulsion among quarks is treated locally, then $c_S^2\rightarrow1$ for constant vector coupling \cite{Blaschke:2022egm} and $c_S^2\rightarrow1/3+0$ in the case when the latter is medium dependent \cite{Ivanytskyi:2022bjc}.
Thus, accounting for the nonlocal character of interaction among quarks is important for reaching the conformal limit of QCD in agreement with the results of perturbative calculations \cite{Kurkela:2009gj,Fraga:2013qra,Gorda:2018gpy,Fernandez:2021jfr}, i.e., as $c_S^2\rightarrow1/3-0$ and $\delta\rightarrow+0$.

\begin{figure}[t]
\includegraphics[width=0.9\columnwidth]{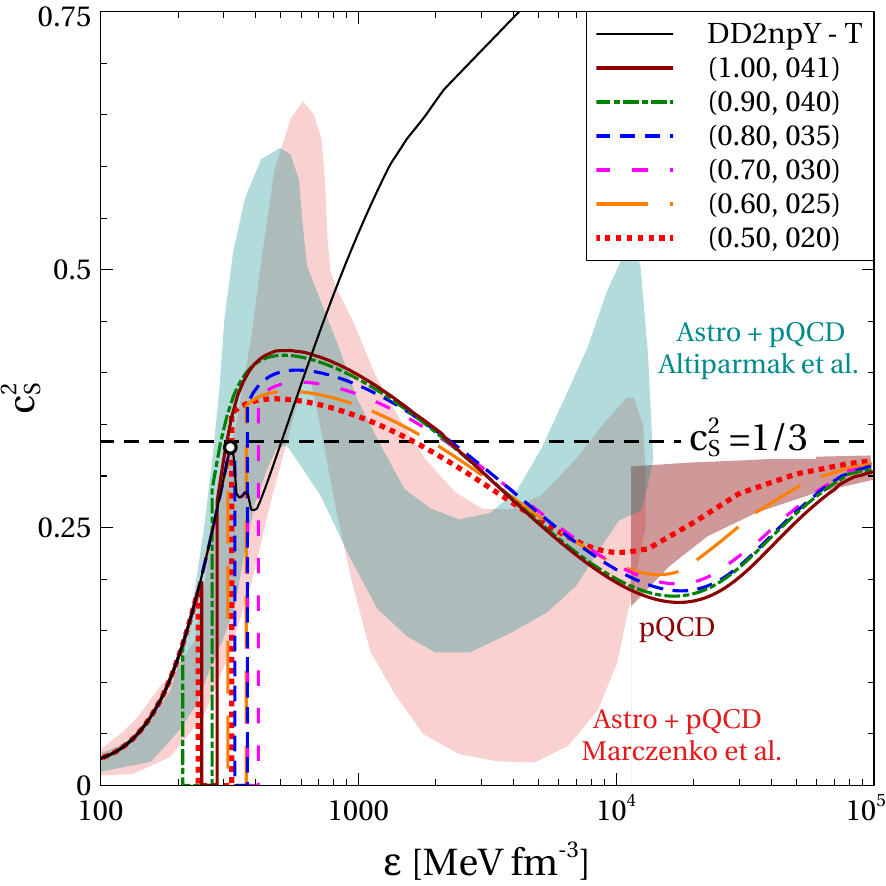}\\
\includegraphics[width=0.9\columnwidth]{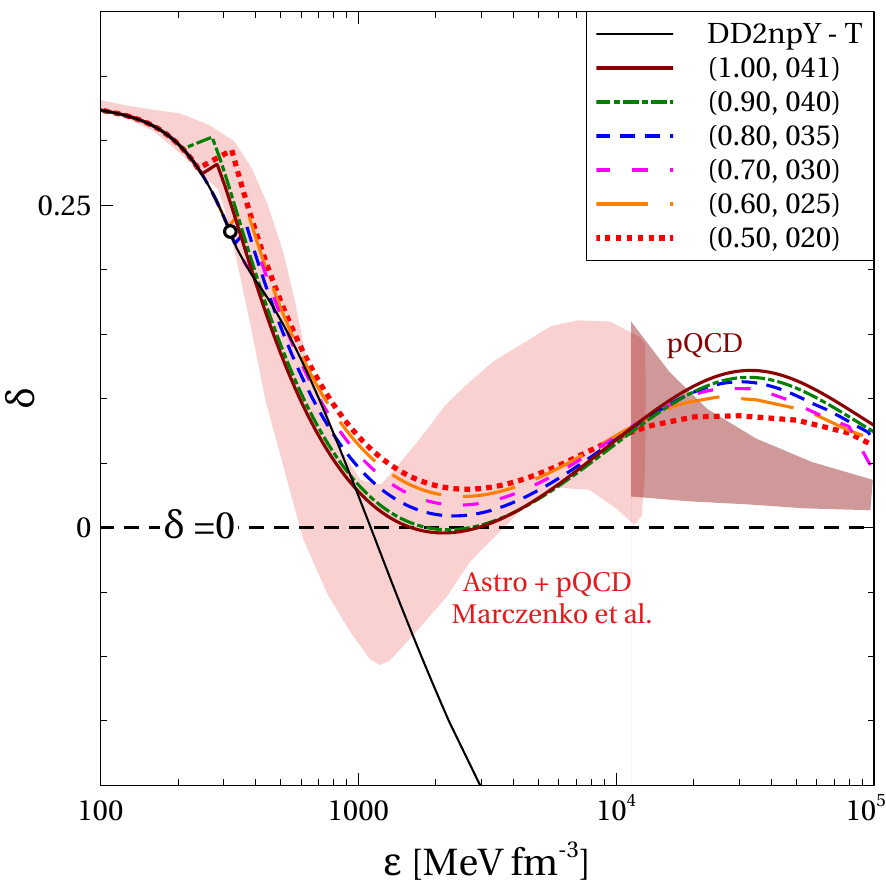}
\caption{Speed of sound $c_S^2$ (upper panel) and dimensionless interaction measure $\delta$ (lower panel) as functions of energy density $\varepsilon$ obtained for the vector and diquark couplings indicated in the legend by the pairs of numbers $(\eta_V,\eta_D)$. 
The empty white circles indicate onset of hyperons in pure hadron matter.
The black dashed lines represent the conformal values $c_S^2=1/3$ and $\delta=0$.
The astrophysical \cite{Altiparmak:2022bke,Marczenko:2022jhl} and perturbative QCD \cite{Fraga:2013qra} constraints represented by the shaded areas are discussed in the text.}
\label{fig1}
\end{figure}

Fig. \ref{fig1} shows the speed of sound and dimensionless interaction measure of several hybrid EOSs obtained by matching the CFLL EOS to the DD2Y-T hadron EOS by means of the Maxwell construction.
At small densities the NS matter is constituted by hadrons with small $c_S^2$ and $\delta\lesssim1/3$.
The mixed quark-hadron phase is characterized by a constant pressure, which leads to vanishing speed of sound and dimensionless interaction measure being hyperbolic in energy density.
Above the mixed phase CFL quarks dominate the NS matter.
In this case, the speed of sound and dimensionless interaction measure exceed their conformal values up to the densities about $2~\rm GeV~fm^{-3}$.
As is shown in Ref. \cite{Ivanytskyi:2024zip} and in Sec. \ref{sec5}, this range of $\varepsilon$ covers the densities accessible in the NS interiors.
This means that the EOS of quark matter, which can exist in NS, is rather stiff and is characterized by superconformal speed of sound.
This superconformal character provides the color-superconducting quark cores with the ability to maintain the existence of NSs with the masses exceeding two solar masses.
As is seen, at the densities typical for NSs the scenario of color-superconducting quark cores in their interiors is consistent with the constraints on $c_S^2$ and $\delta$ extracted from the observational data \cite{Altiparmak:2022bke,Marczenko:2022jhl}.
It is also seen in Fig. \ref{fig1} that $c_S^2$ and $\delta$ of the CFL quark matter simultaneously attain nearly conformal values at the energy density $\varepsilon\simeq2~\rm GeV~fm^{-3}$.
However, as is explained in Ref. \cite{Ivanytskyi:2024zip}, this does not necessarily signals about approximately conformal behavior of the CFL quark matter.
Above $\varepsilon\simeq2~\rm GeV~fm^{-3}$ $c_S^2$ becomes subconformal, while $\delta$ exceeds its conformal value.
Fig. \ref{fig1} shows that the CFL quark matter asymptotically reaches the conformal limit of QCD. 
It is also in line with the asymptotes of $c_S^2$ and $\delta$ found in Ref. \cite{Ivanytskyi:2024zip}, which agree with the results of perturbative QCD \cite{Fraga:2013qra}.

\section{The set of EOS}
\label{sec4}

\begin{figure}[t]
    \includegraphics[width=\columnwidth]{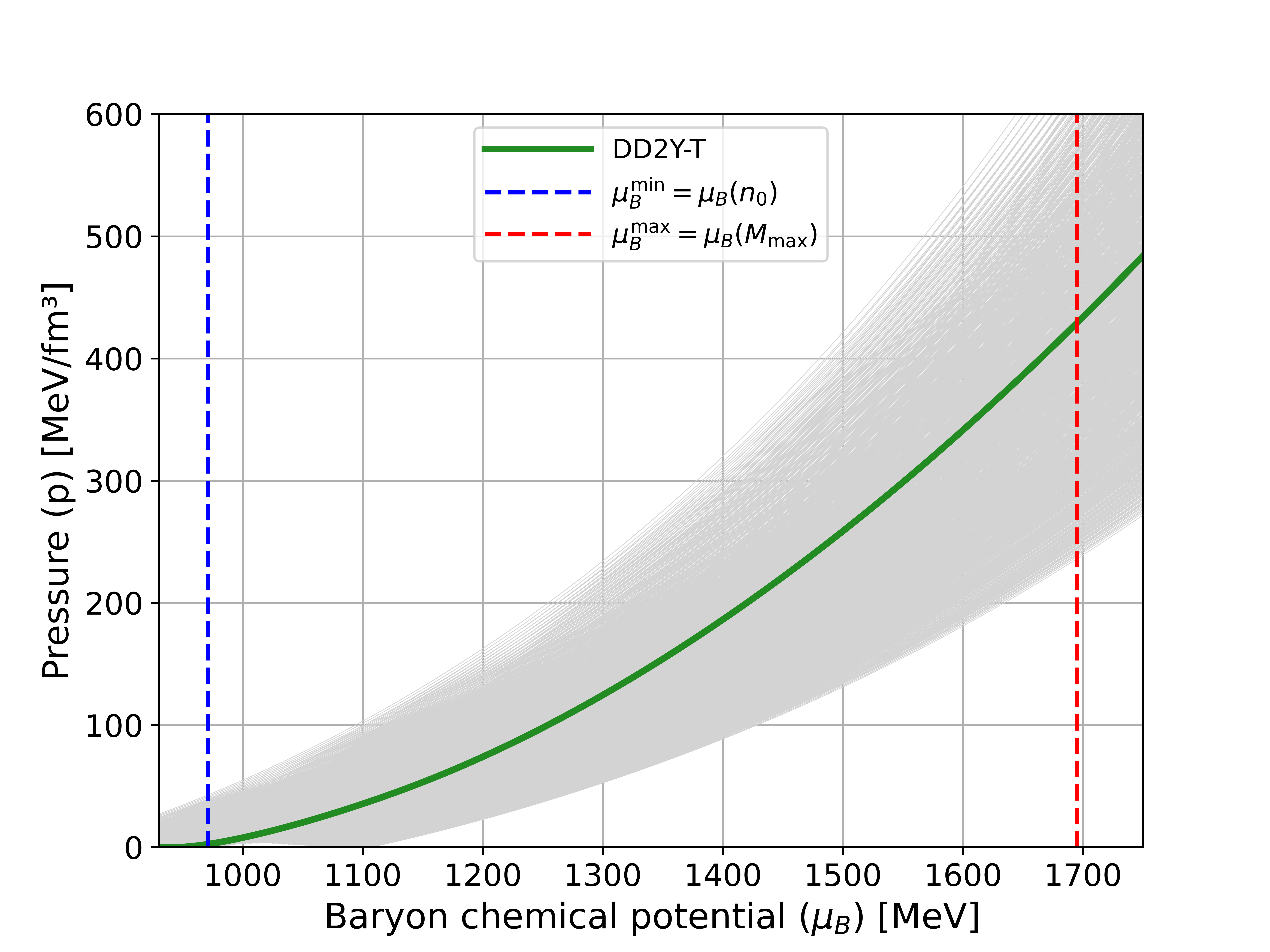}
    \caption{Pressure $p$ as a function of baryon chemical potential $\mu_B$ of all the generated quark EOSs (thin gray curves) and the DD2Y-T hadronic EOS (bold green curve). The vertical dashed lines indicate the minimal $\mu_B^{\rm min}$ and maximal $\mu_B^{\rm max}$ allowed values of the baryon chemical potential at the quark-hadron phase transition (see text for details).}
    \label{fig:p_vs_mu}
\end{figure}

Performing a reliable Bayesian analysis of the available observational data on NSs requires generating a full set of EOS of NSs.
To generate this set we first produced a collection of $31\times36$ quark EOSs labeled by the values of the vector and diquark couplings of the microscopic Lagrangian.
These couplings are varied in the wide ranges of $\eta_D=0.20(0.01)0.50$ and $\eta_V=0.50(0.02)1.20$.
This ensures the generality and inclusive character of our analysis, which accounts for many physically motivated realizations of the asymptotically conformal quark EOS with vector repulsion and CFL color superconductivity.
The mentioned range of the diquark coupling also respects the requirement that the vacuum state is stable against formation of color superconductivity, i.e., $\eta_D\le0.765$ \cite{Ivanytskyi:2024zip}.
The gray curves in Fig.~\ref{fig:p_vs_mu} demonstrate the generated quark EOSs compared to the baseline hadronic DD2Y-T EOS \cite{Shahrbaf:2022upc} depicted by the green curve. 

\begin{figure}[t]
    \includegraphics[width=\columnwidth]{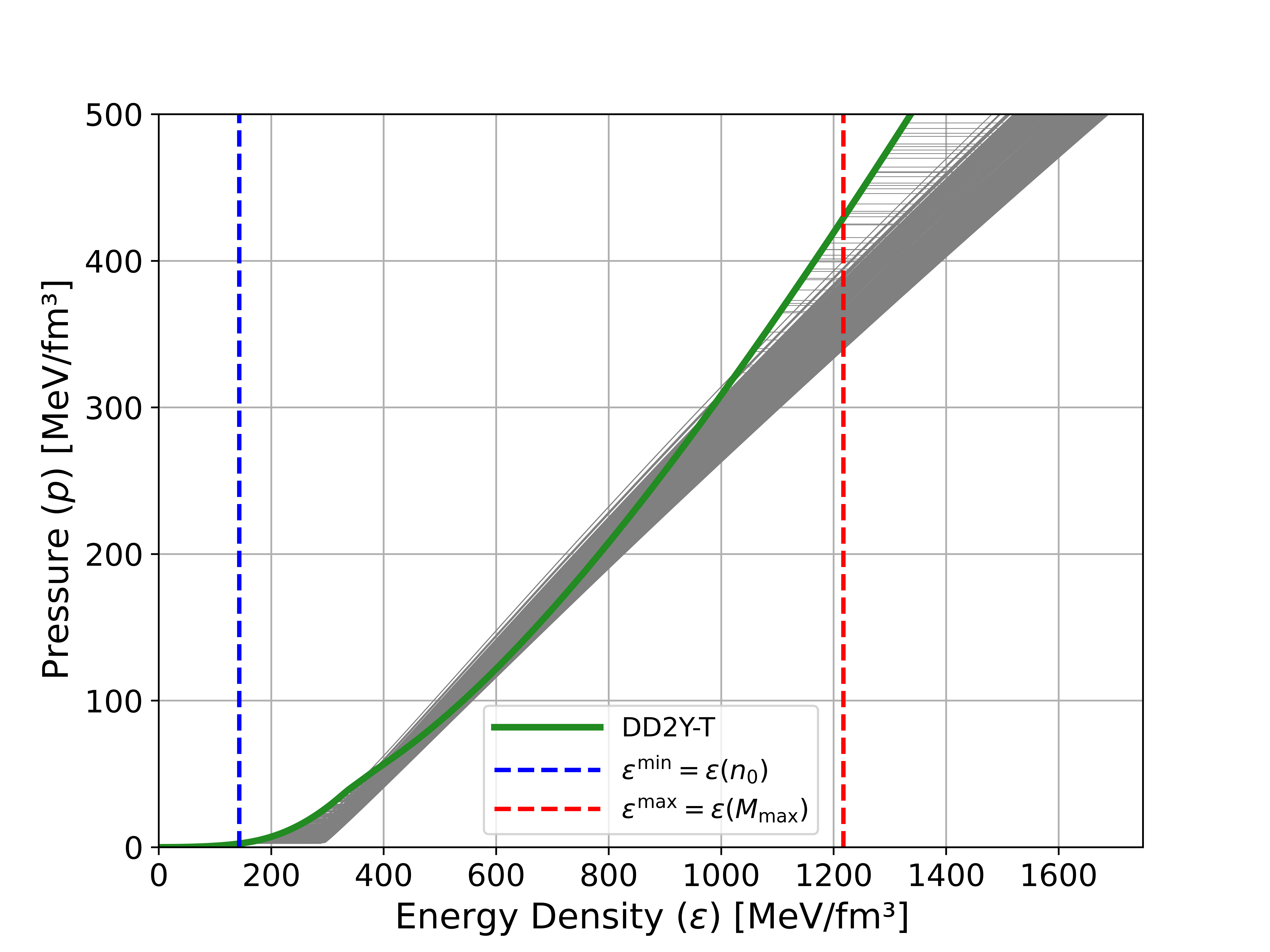}
    \caption{Pressure $p$ as a function of energy density $\varepsilon$ of the hybrid EOSs with a proper Maxwell crossing and the transition at baryon chemical potentials $\mu_B$ above $\mu_B^{\rm min}$ and $\mu_B^{\rm max}$ depicted by blue and red dashed lines, respectively, (thin gray curves). The bold green curve represents the DD2Y-T hadronic EoS.}
    \label{fig:hybrid_p_vs_eps}
\end{figure}

Later, it is checked which of the generated quark EOSs provide crossings with the hadronic DD2Y-T EOS in the plane of pressure $p$ versus baryon chemical potential $\mu_B$.
This corresponds to the Maxwell construction of a first-order quark-hadron transition.
It is obvious that many quark EOSs do not provide such crossing.
They are not considered in the further analysis.
Some of the quark EOSs have Maxwell crossings so that the slopes of their pressures as a function of baryon chemical potential is smaller than the corresponding slope of the hadron EOS.
In this case the baryon density experiences a negative jump across the phase transition if hadron matter converts to the quark one with increasing the chemical potential.
For these quark EOSs a positive density jump across the phase transition is possible only if quark matter is favored at small chemical potentials, where it has a higher pressure, while the hadronic matter is the dominating phase at high chemical potentials.
This picture contradicts to the commonly accepted phenomenology of QCD and, therefore, the corresponding quark EOSs are also excluded from the analysis.
The remaining 307 quark EOSs provide a positive density jump at the Maxwell crossing with the DD2Y-T EOS.
Each of these EOSs has a specific onset density of quark matter.
We require it to exceed one nuclear saturation density, which corresponds to the minimal value of the onset chemical potential $\mu_B^{\rm min}=971$.
We also restrict the onset density to the values, which are below the central density in the heaviest NS with purely hadronic DD2Y-T EOS.
This is equivalent to requiring that the baryon chemical potential does not exceed the maximal onset value $\mu_B^{\rm max}=1695$ MeV.
The later requirement is necessary for the TOV-stability of the quark branches of the mass-radius relations of the NSs with quark cores.

Fig. \ref{fig:hybrid_p_vs_eps} shows the EOSs consistent with the above requirements in the plane of pressure versus energy density.
As is seen, at $\varepsilon\simeq400-1000~{\rm MeV/fm^3}$ some hybrid EOSs are stiffer than the purely hadronic one.
These EOSs allow reaching the maximum NS masses about $2.2~{\rm M}_\odot$, which is not possible with the DD2Y-T hadronic EOS.

\begin{figure}[t]
    \includegraphics[width=\columnwidth]{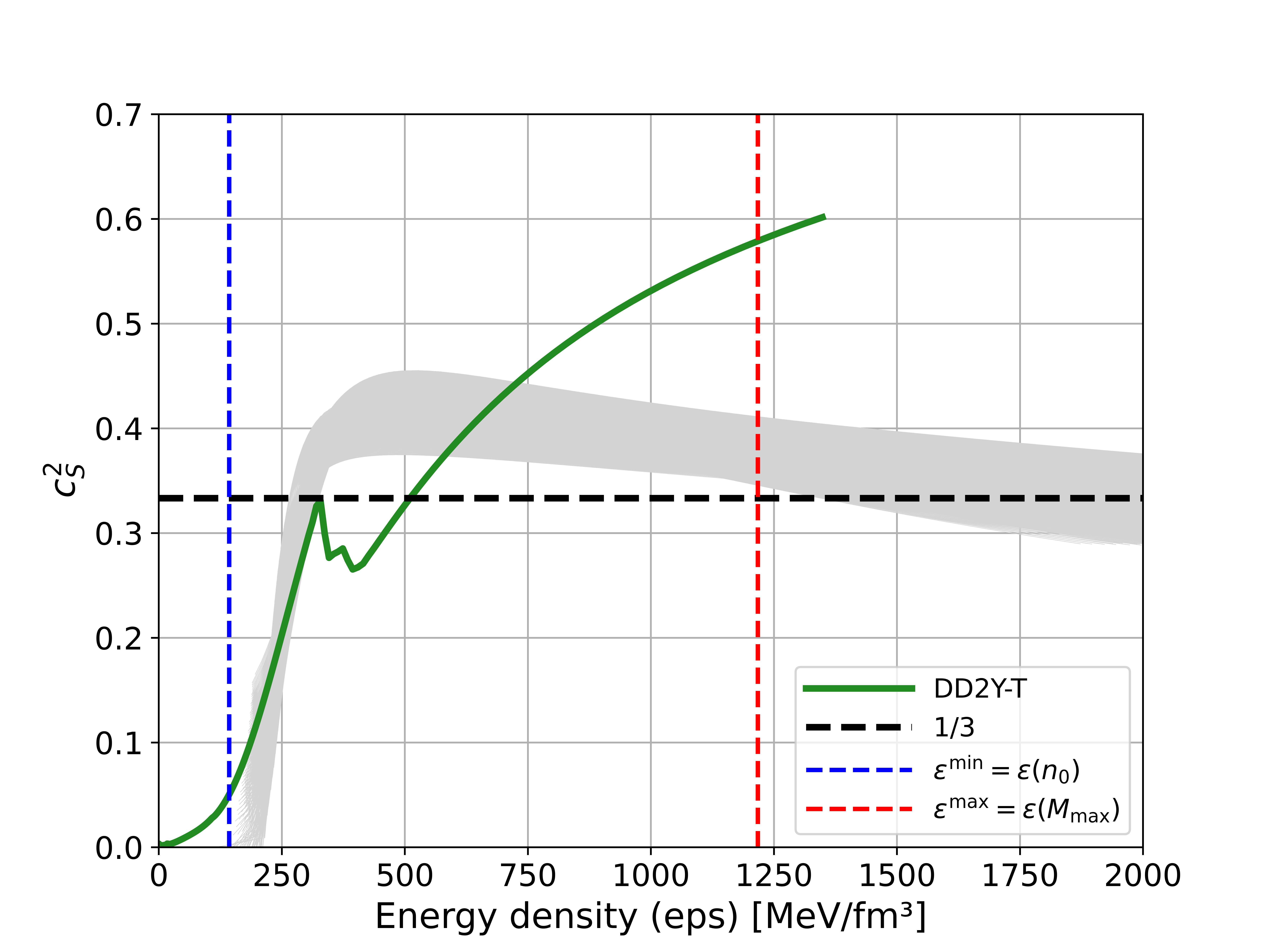}
    \caption{Squared speed of sound $c_S^2$ as a function of energy density $\varepsilon$ of the EOSs shown in Fig. \ref{fig:p_vs_mu} along with the same minimal and maximal allowed values of the energy density of the onset of quark matter (see text for details).}
    \label{fig:phases_macro}
\end{figure}

The squared speed of sound of the hybrid EOSs is shown in Fig. \ref{fig:phases_macro}.
The general structure discussed in Sec. \ref{sec2} holds.
It is also seen that in the density range typical for the NS interiors quark branch of the hybrid EOS has superconformal speed of sound.
This provides the NS matter with the stiffness required to maintain the existence of the objects heavier than two solar masses.

\begin{figure}[t]
    \centering
    \includegraphics[width=\columnwidth]{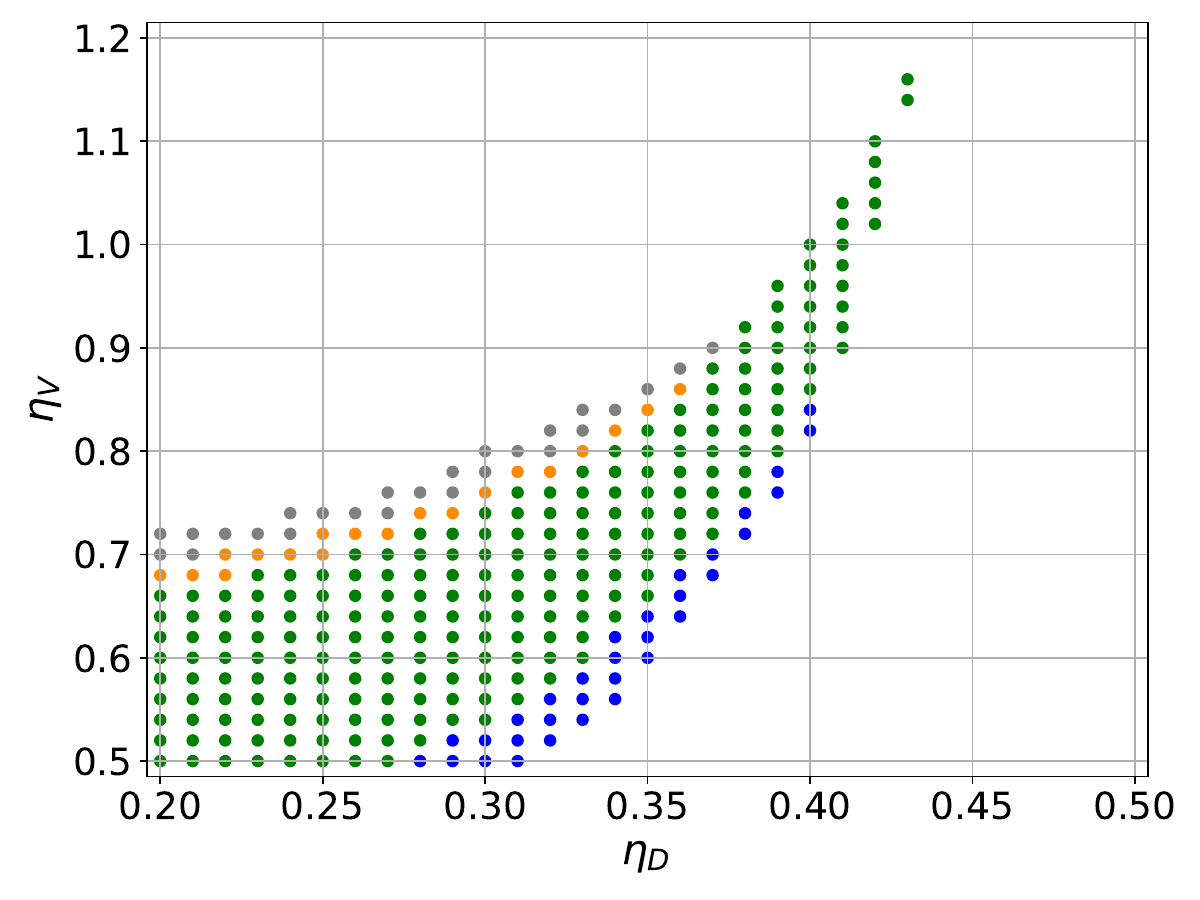}
    \caption{The region in the plane of the vector $\eta_V$ versus diquark $\eta_D$ couplings providing the existence of hybrid quark-hadron EOSs with a positive density jump across the phase transition. 
    The green and blue dots represent the EOSs, which support the TOV-stability of the quark branches of the NS mass-radius relations, while the orange and gray dots do not provide such stability.
    The blue dots correspond to the EOSs fulfilling the Seidov criterion of the existence of twin NSs. The gray dots show
    the EOSs with a phase transition at $\mu_B>\mu_B^{\rm max}$, when the onset density exceeds the central density in the maximum-mass hadronic configurations.
    }
    \label{fig:phases_unconstrained}
\end{figure}

The 269 combinations of $\eta_V$ and $\eta_D$, which provide a Maxwell crossing with a positive density jump at $\mu_B^{\rm min}\le\mu_B\le\mu_B^{\rm max}$ are shown in Fig. \ref{fig:phases_unconstrained}.
As is seen, the corresponding region does not include all the possible values of the vector and diquark couplings.
In general, the higher the vector coupling the higher the chemical potential of the quark-hadron transition, while changing the diquark coupling causes the opposite effect.
Increasing $\eta_D$ at a given value of $\eta_V$ shifts the phase transition toward small chemical potentials until it reaches $\mu^{\rm min}_B$.
This corresponds to the lower right boundary of the region shown in Fig. \ref{fig:phases_unconstrained}.
The hybrid EOSs represented by the blue dots located along this boundary fulfill the Seidov criterion of gravitational instability due to a large energy density jump across the phase transition \cite{1971SvA15347S}.
If at higher densities after the transition the gravitational stability is regained due to a significant stiffening of quark EOS, then a new stable branch of hybrid NS appears, which is disconnected from the purely hadronic branch of NS mass-radius relations.
These branches include pairs of NSs of the same mass but different radii, so-called twin stars \cite{Blaschke:2019tbh,Goncalves:2022phg,Chanlaridis:2024rov}. 
The hybrid star branches of the mass-radius relations of the NSs with these EOSs are TOV-stable.
The green dots in Fig. \ref{fig:phases_unconstrained} represent the EOSs, which also provide the TOV-stability of their quark branches but break the Seidov criterion.
Decreasing $\eta_D$ at constant $\eta_V$ shifts the phase transition toward higher chemical potentials.
At high values of the vector coupling this happens until the phase transition disappears at $\mu_B<\mu^{\rm max}_B$ since the quark EOS gets too stiff.
This corresponds to the $\eta_V\gtrsim0.87$ part of the upper left boundary of the region shown in Fig. \ref{fig:phases_unconstrained}.
At $\eta_V\lesssim0.87$ this boundary is defined by higher chemical potentials.
The orange and gray dots located along this boundary represent the quark EOSs, which provide a proper Maxwell crossing but generate mass-radius relations with unstable quark branches. 
The corresponding regions of the $\eta_V-\eta_D$ plane are separated by the condition $\mu_B=\mu^{\rm max}_B$.
In other words a Maxwell crossing with the positive density jump happens at $\mu_B<\mu^{\rm max}_B$ for the orange dots and at $\mu_B>\mu^{\rm max}_B$ for the gray ones.
The corresponding EOSs are not used in the further analysis.

Thus, only 269 quark EOSs, which provide a Maxwell crossing with a positive density jump above the nuclear saturation density and the TOV-stability of the quark branches of the mass-radius relations of the corresponding NSs are used to perform the Bayesian analysis of the available observational data.
In Fig. \ref{fig:phases_unconstrained} these 250 EOSs are represented by the green and blue dots.
The stiffness of quark EOS increases when moving from the lower left to the upper right part of the corresponding region in the $\eta_V-\eta_D$ plane.

\section{The Bayesian framework}
\label{sec5}

To infer the posterior probability distribution of the EOS parameters from observational data, we employ a Bayesian approach. 
According to the Bayes theorem, the posterior probability of describing the data set $\mathcal{D}$ with the EOS constructed at given values of couplings $\eta_V$ and $\eta_D$ takes the form of
\begin{equation}
  p((\eta_D,\eta_V)\,|\,\mathcal{D}) = 
  \frac{p(\mathcal{D}\,|\,(\eta_D,\eta_V)) \, p(\eta_D,\eta_V)}{p(\mathcal{D})},
\end{equation}
where $p(\mathcal{D}\,|\,(\eta_D,\eta_V))$ denotes the likelihood, $p(\eta_D,\eta_V)$ represents the prior probability distribution, and $p(\mathcal{D})$ is evidence. 
We adopt the uniform prior
\begin{equation}
  p(\eta_D,\eta_V) = \frac{1}{|(\eta_D,\eta_V)|} = \frac{1}{N}.
\end{equation}

The total likelihood is obtained as a product over all 
independent observational data $D_\alpha$,
\begin{equation}
  p(\mathcal{D}\,|\,(\eta_D,\eta_V)) = 
  \prod_{\alpha}^N p(D_\alpha\,|\,(\eta_D,\eta_V)).
\end{equation}

The likelihood associated with the lower bound on the 
maximum mass is modeled by a normal cumulative distribution 
function $F_\mathcal{N}$,
\begin{equation}
  p(D_{M}\,|\,(\eta_D,\!\eta_V)) = 
  F_\mathcal{N}\!\left(M_{\max}(\eta_D,\!\eta_V);\,
  \mu_M,\sigma_M\right),
\end{equation}
with $\mu_M$ and $\sigma_M$ denoting the mean and 
uncertainty of the constraint taken from 
Refs.~\cite{Romani:2022jhd}.

For mass--radius data the likelihood is evaluated by integration over the mass-radius curve obtained for the corresponding EOS
\begin{equation}
  p(D_{M\!R}\,|\,(\eta_D,\!\eta_V)) = 
  \int\limits_{M\!R}
  f_{M\!R}\!\left(M(\eta_D,\!\eta_V),\,R(M)\right).
\end{equation}
When disconnected branches of the mass-radius relation occur, this integral must be evaluated separately over each branch and the contributions summed. 
The functions $f_{MR}^{(i)}$ are constructed via Kernel Density 
Estimation (KDE)~\cite{Chacon2020-ql} from the Zenodo data for 
PSR~J0030+0451~\cite{miller:2019:3473466}, 
PSR~J0740+6620~\cite{dittmann:2024:10215108}, 
PSR~J0437--4715~\cite{choudhury:2024:13766753}, 
PSR J0614--3329~\cite{mauviard_2025_15603406}
and HESS~J1731-347~\cite{doroshenko:2023:8232233}.

The gravitational wave (GW) likelihood is defined in an analogous way, i.e., through the integral over the relation between the tidal deformabilities of two NSs with the proper chirp mass
\begin{equation}
  p(D_{GW}\,|\,(\eta_D,\!\eta_V)) = 
  \int\limits_{\Lambda_1\!\Lambda_2}
  f_{GW}\!\left(\Lambda_1(\eta_D,\!\eta_V), \Lambda_2(\Lambda_1)\right),
\end{equation}
where $f_{GW}$ is based on the GW170817 data~\cite{ligoLIGOP1800115v12GW170817}.

The evidence follows from marginalization over all possible parameter values,
\begin{equation}
  p(\mathcal{D}) = \sum_{(\eta_D,\!\eta_V)} 
  p(\mathcal{D}\,|\,(\eta_D,\!\eta_V))\, p(\eta_D,\!\eta_V).
\end{equation}

We perform the Bayesian analysis (BA) under two different scenarios. 
The first one, denoted as the \textit{basic set}, includes all currently available NICER mass–radius constraints:
\begin{itemize}
    \item $M = 1.44_{-0.14}^{+0.15}~M_{\odot}$ and $R = 13.02_{-1.06}^{+1.24}~\mathrm{km}$ for PSR~J0030+0451~\cite{Miller:2019cac},
    \item $M = 2.08_{-0.07}^{+0.07}~M_{\odot}$ and $R = 12.92_{-1.13}^{+2.09}~\mathrm{km}$ for PSR~J0740+6620~\cite{Dittmann:2024mbo}, which refines the earlier NICER measurement of~\cite{Miller:2021qha},
    \item $M = 1.418_{-0.037}^{+0.037}~M_{\odot}$ and $R = 11.36_{-0.63}^{+0.95}~\mathrm{km}$ for PSR~J0437--4715~\cite{Choudhury:2024xbk}, and
    \item $M = 1.44_{-0.07}^{+0.06}~M_{\odot}$ and $R = 10.29_{-0.86}^{+1.01}~\mathrm{km}$ for PSR~J0614--3329~\cite{Mauviard:2025dmd}.
\end{itemize}
In addition, this set incorporates the tidal deformability constraint from GW170817~\cite{LIGOScientific:2018cki}.

The second scenario, referred to as the \textit{full set}, extends the basic set by including two further measurements:
\begin{itemize}
    \item $M = 2.35_{-0.17}^{+0.17}\times f(\Omega)~M_{\odot}$ for the ``black widow'' pulsar PSR~J0952--0607~\cite{Romani:2022jhd}\footnote{Here we apply a correction factor $f(\Omega)=[1+0.2\,(\Omega/\Omega_K)^2]^{-1}$ because the pulsar is one of the fastest spinning pulsars at about half of its Kepler frequency, $\Omega \sim 0.5\, \Omega_K$, so that the well known $20\%$ increase of the mass for a pulsar spinning at the Kepler frequency $\Omega_K$ \cite{Haensel:2007yy,Falcke:2013xpa,Breu:2016ufb,Largani:2021hjo}
    together with the quadratic scaling amounts to a down correction of the mass constraint by $5\%$ when it is used as a constraint for mass-radius relations on nonrotating stars from the solutions of the TOV equations.}, which provides a lower limit on the maximum mass, and
    \item $M = 0.77_{-0.17}^{+0.20}~M_{\odot}$ and $R = 10.04_{-0.78}^{+0.86}~\mathrm{km}$ for HESS~J1731--347~\cite{Doroshenko:2022nwp}.
\end{itemize}
We note that historically Bayesian analyses have tried to implement as many precise mass and radius constraints as were available to infer the most likely EOS of NS matter. Following this strategy a problem arises when a mass constraint like the one for the $2\rm M_\odot$ pulsar J0348+0432 \cite{Antoniadis:2013pzd} with unconstrained radii is used together with the NICER constraint for PSR J0740+6620 that excludes too compact high-mass stars with $R_{2.0} < 11.79$ km. We argue that once the NICER constraint is applied, the mass constraint for PSR J0348+0432 should not be used. Otherwise, the selective power of the NICER constraint of the $R_{2.0}$ measurement will be eliminated.
For a detailed discussion and the effect on the credibility regions in Fig. \ref{fig:BA2}, see Appendix \ref{appB}.

\begin{figure}[t]
    \includegraphics[width=\columnwidth]{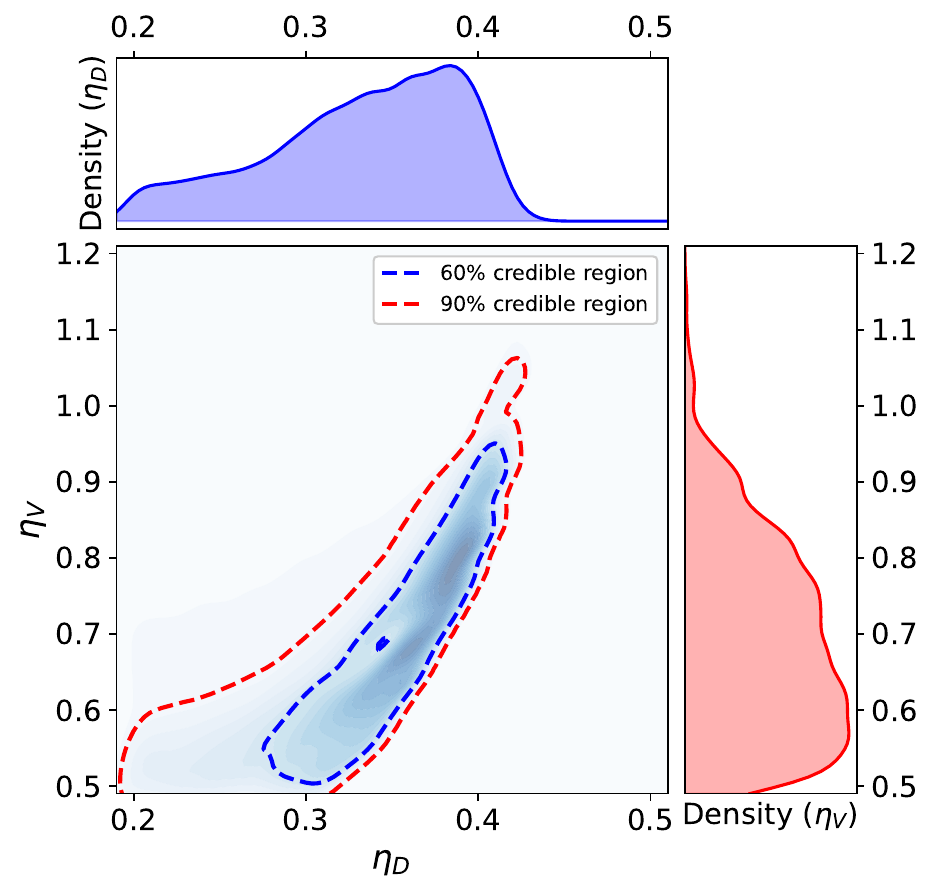}
    \includegraphics[width=\columnwidth]{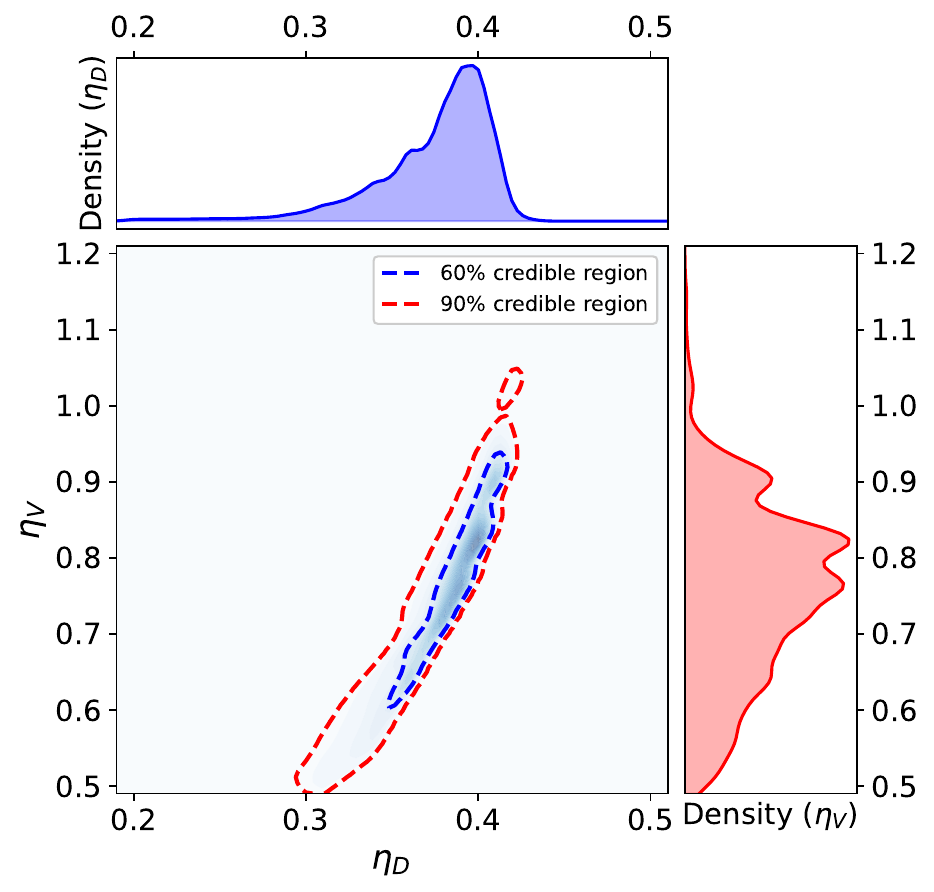}
   \caption{The posterior probability as a function of the vector $\eta_V$ and diquark $\eta_D$ couplings obtained from the Bayesian analysis of the basic (upper panel) and full (lower panel) sets of the observational data
   }
  \label{fig:BA}
\end{figure}

\section{Results and discussion}
\label{sec6}

Fig. \ref{fig:BA} shows the posterior probability as a function of the vector and diquark couplings evaluated for the basic and full sets of the observational data.
The probability peaks around $\eta_V\simeq0.64$ and $\eta_D\simeq0.36$ for the basic set.
Not too high values of the most probable value of the vector coupling correspond not too stiff EOSs.
This does not allow the maximum NS masses to reach the high value of the black widow pulsar PSR J0952-0607 \cite{Romani:2022jhd}.
Therefore, in the full set case the maximum of the posterior probability is located at higher vector coupling around $\eta_V\simeq0.82$ and $\eta_D\simeq0.40$.
In this case, the most probable diquark coupling is also higher than for the basic set. 
This is due to the fact that the data from GW170817 ~\cite{ligoLIGOP1800115v12GW170817} and HESS~J1731-347~\cite{doroshenko:2023:8232233} prefer relatively small radii of the intermediate mass NSs, which can be provided by phase transition before $1\rm M_\odot$ requiring higher values of the diquark coupling if the vector one is also high.
Based on this we conclude that the most probable and physically interesting values of the vector and diquark couplings are $\eta_V\simeq0.82$ and $\eta_D\simeq0.40$.

\begin{figure}[t]
    \includegraphics[width=\columnwidth]{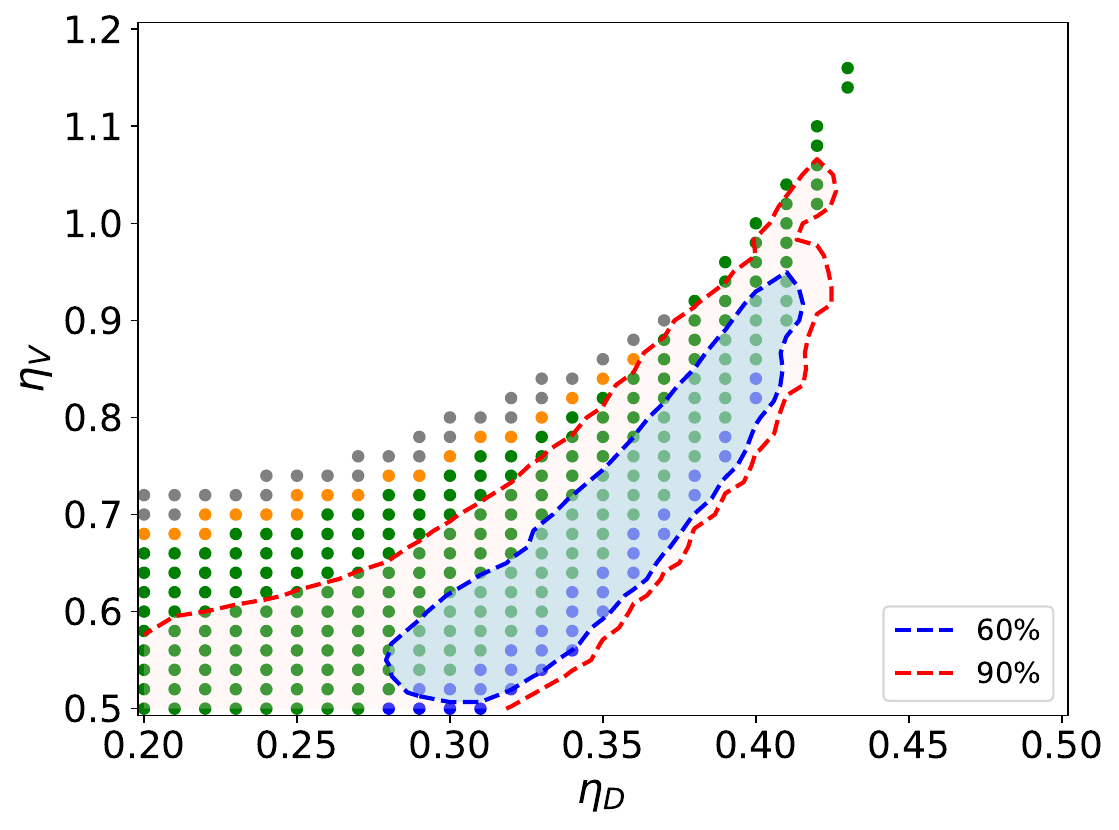}
    \includegraphics[width=\columnwidth]{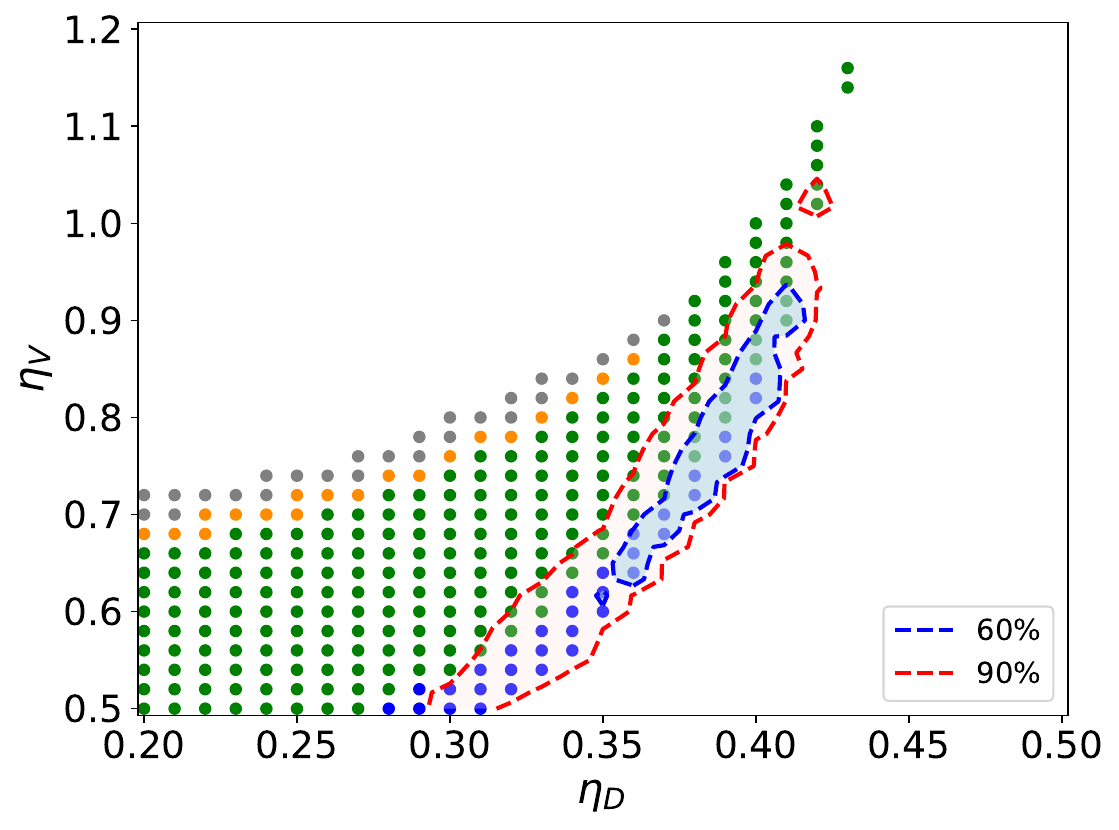}
    \caption{The same as in Fig. \ref{fig:phases_unconstrained} but with overlaid 60\% and 90\% credibility contours obtained for the basic (upper panel) and full (lower panel) sets of constraints from the observational data depicted in Fig. \ref{fig:BA}.
    }
  \label{fig:BA2}
\end{figure}

Fig. \ref{fig:BA} also shows the 90\% and 60\% confidence regions.
As is seen, analyzing the full data set shifts these regions toward higher $\eta_V$ and $\eta_D$, and narrows them compared to the case of the basic set.
It is notable that compared to the recent results of the physically-educated Bayesian analysis \cite{Ayriyan:2024zfw} performed with the two-flavor NJL model with local vector repulsion \cite{Contrera:2022tqh}, the 60\% and 90\% confidence regions obtained in this work do not include small values of the vector coupling $\eta_V\lesssim0.5$ and $\eta_V\lesssim0.6$ corresponding to the basic and full sets, respectively.
Such small values are disfavored by the observational data since they lead to an excessive softness of quark EOSs that cannot be compensated by the vector repulsion, which in the nonlocal case is subdominant at high densities \cite{Ivanytskyi:2024zip}. 

Fig. \ref{fig:BA2} demonstrates the 60\% and 90\% confidence contours overplayed on the $\eta_V-\eta_D$ region showing the combinations of the couplings, which provide the existence of hybrid quark-hadron EOSs with a positive density jump across the phase
transition.
As is seen, the couplings leading to the quark onset density above the central density in the heaviest purely hadronic NS (gray dots in Fig. \ref{fig:BA2}) are beyond these contours.
In other words, most of the hybrid EOSs provide significantly better agreement with the observational data on NSs than the benchmark DD2Y-T EOS.
This is a statistical evidence in favor of the scenario of NSs with quark cores compared to the purely hadronic scenario.
The robustness of this evidence can be quantified by comparing the value of the posterior probability at its global maximum to the probability of the hadronic DD2Y-T EOS. 
The corresponding ratio is 18.73 and 276.15 for the basic and full sets of the observational data, respectively.

\begin{figure}[t]
    \includegraphics[width=1.1\columnwidth]{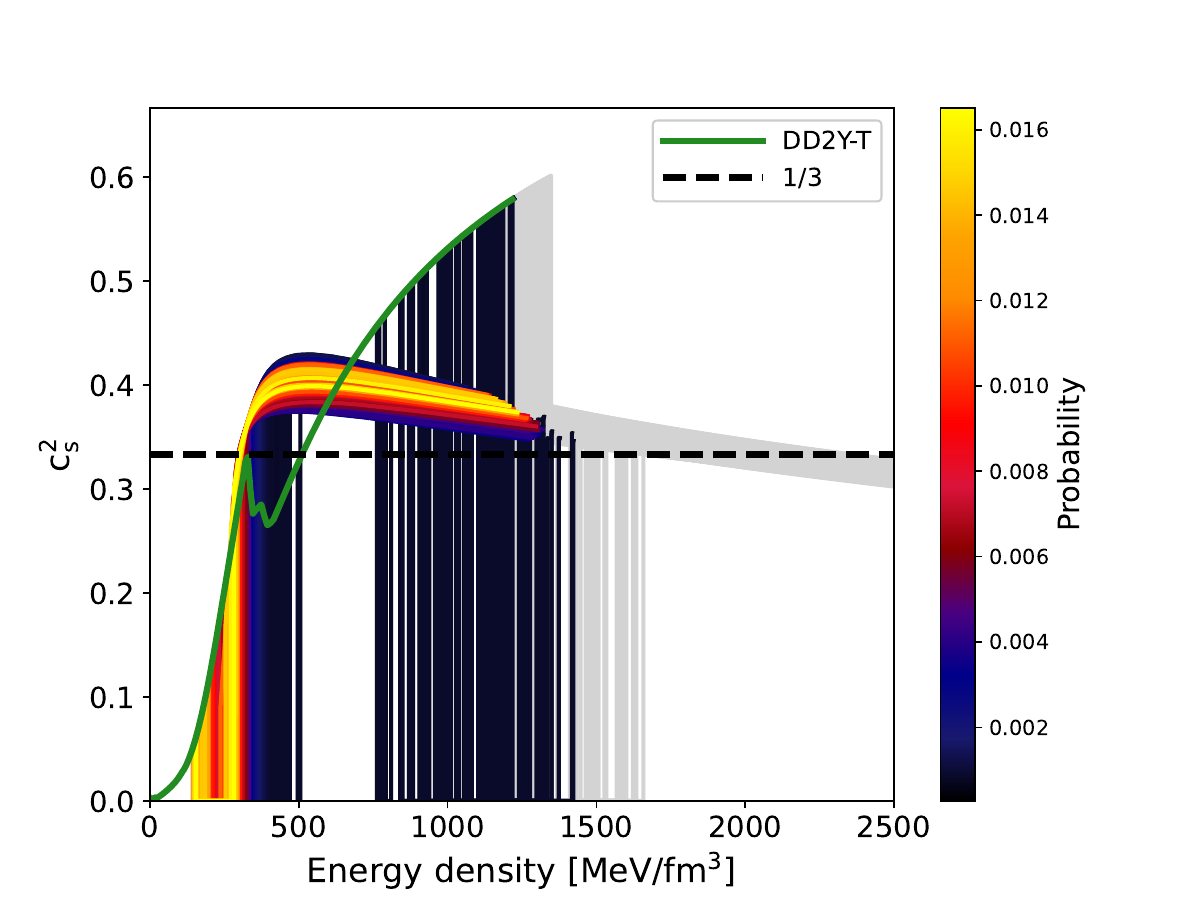}
    \caption{Speed of sound squared $c_S^2$ of the NS matter as a function of energy density $\varepsilon$.
    The color map corresponds to the posterior probability evaluated using the full set of the observational data.
    The gray parts of the curves correspond to the densities beyond the central density in the centers of the heaviest NSs.}
    \label{fig:cs2_eps}
\end{figure}

Fig. \ref{fig:cs2_eps} shows the squared speed of sound of the EOSs used for the Bayesian analysis of the full set of the observational data.
Note, the results corresponding to the basic set are not shown since visually they are almost identical.
The first important conclusion is that the most probable EOSs have a first order phase transition at the energy density $\varepsilon=150-300~\rm MeV/fm^3$.
This range agrees well with the recent results of the Bayesian analysis with the two-flavor chiral quark model with local vector repulsion \cite{Ayriyan:2024zfw}.
The values of $\varepsilon$ within this range are rather small and correspond to $1-2$ nuclear saturation densities.
Similar small values of the onset density of deconfinement have also been reported in Refs. \cite{Gartlein:2023vif,Gartlein:2024cbj}.
They should not be perceived as a contradiction to the fact that the experiments on relativistic HIC do not indicate a phase transition up to the densities about 5 nuclear saturation ones \cite{Danielewicz:2002pu,Iancu:2012xa,Shuryak:2014zxa,Busza:2018rrf}.
The resolution of this apparent paradox is related to a strong dependence of the onset density of quark matter on isospin asymmetry \cite{Sagert:2008ka}.
The very recent analysis of this dependence indicates that the onset density of deconfinement in the symmetric matter created in HIC can exceed the onset density in strongly asymmetric NS matter by a factor $3-4$ \cite{Panasiuk:2025}. 

It is also seen from Fig. \ref{fig:cs2_eps} that in the most probable scenarios, the NS matter becomes superconformal immediately after the phase transition. 
The corresponding speed of sound reaches its maximal value $c_S^2\simeq0.40$ at the energy density $\varepsilon\simeq500~\rm MeV/fm^3$.
Later, the squared speed of sound gradually decreases reaching the value $c_S^2\simeq0.38$ at the density $\varepsilon\simeq1250~\rm MeV/fm^3$, which is the central density of the heaviest NSs.
Thus, the behavior of the NS matter is likely to be nonconformal since its squared speed of sound significantly exceed the conformal value $c_S^2=1/3$.

\begin{figure}[t]
    \includegraphics[width=0.94\columnwidth]{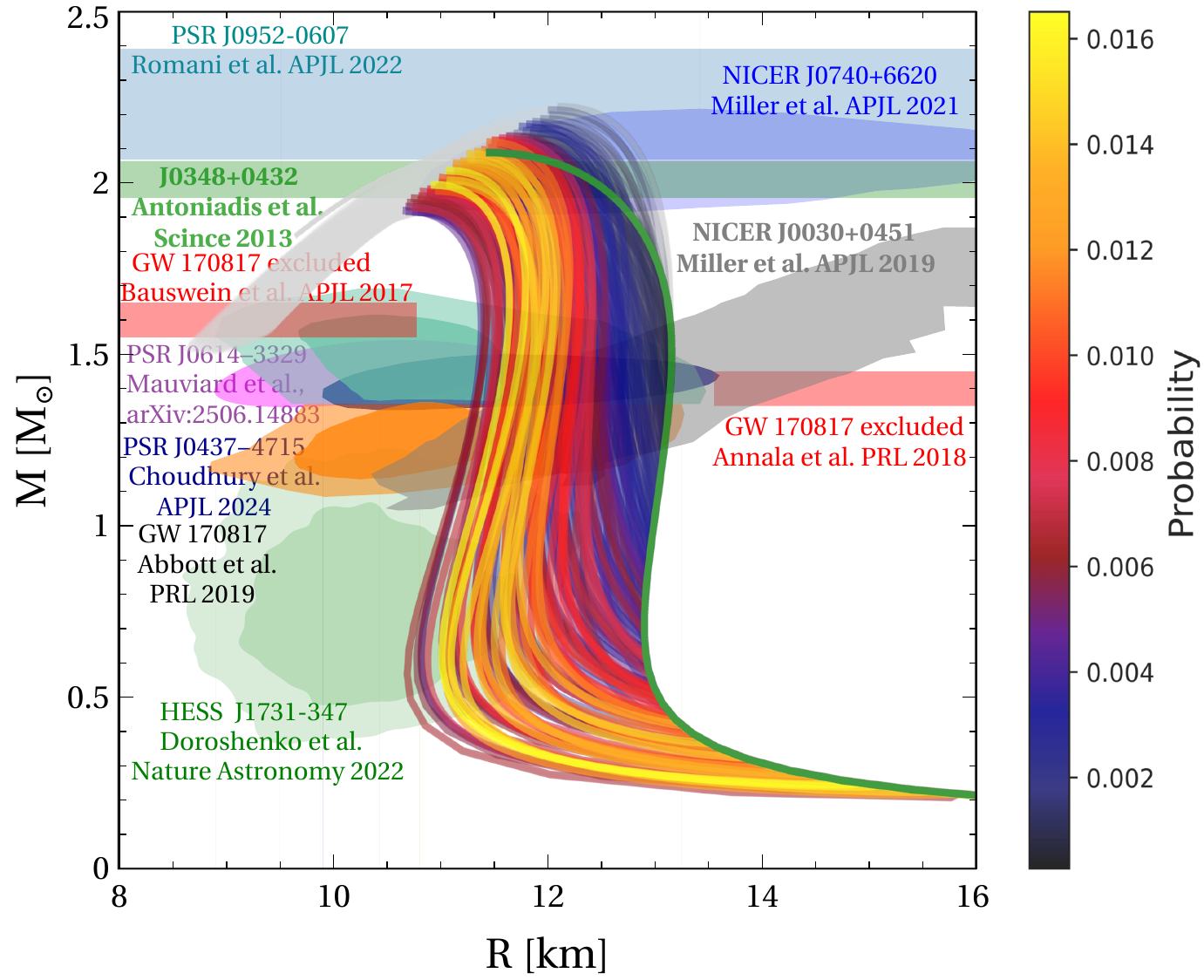}
    \includegraphics[width=1.05\columnwidth]{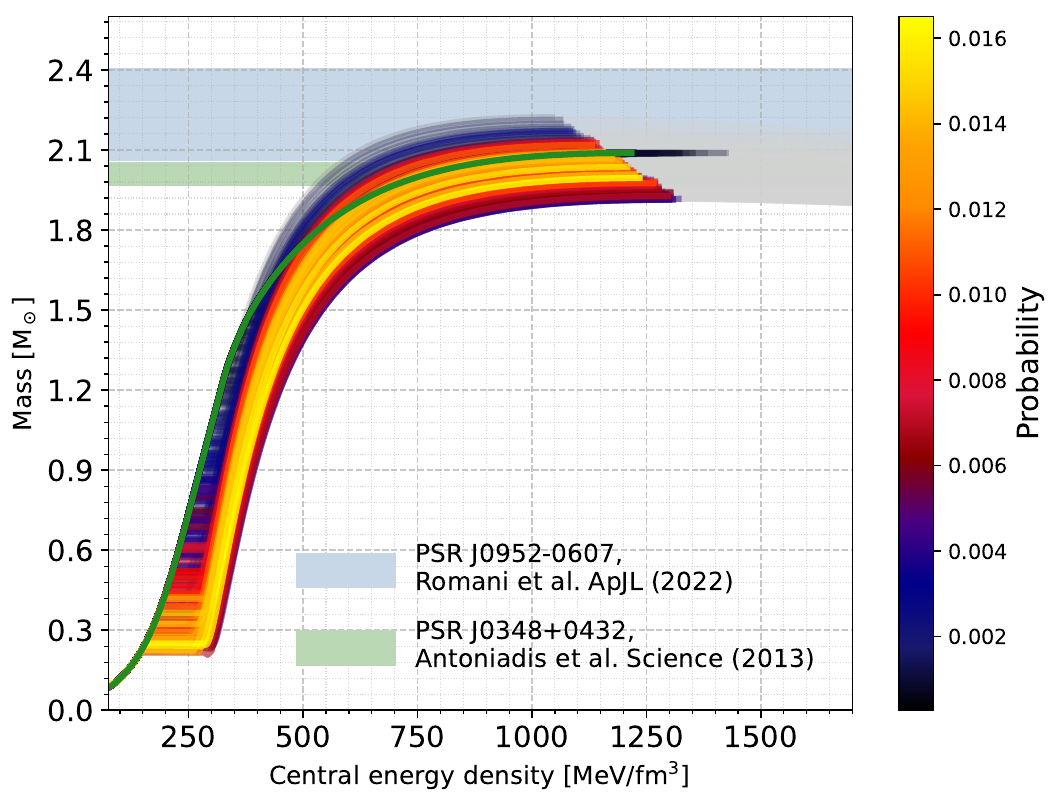}
    \caption{Dependence of the mass $\rm M$ of NS on their radius $\rm R$ (upper panel) and central energy density $\varepsilon_c$ (lower panel).
    The color map corresponds to the posterior probability evaluated using the full set of the observational data.
    The gray parts of the curves correspond to unstable configurations which have central energy densities exceeding the value for the maximum mass configuration.
    The shaded areas represent the observational constraints from Refs. \cite{Antoniadis:2013pzd,Fonseca:2021wxt,Miller:2021qha,Riley:2021pdl,Miller:2019cac,Riley:2019yda,Raaijmakers:2019qny,LIGOScientific:2018cki,Bauswein:2017vtn,Annala:2017llu}.}
    \label{fig:mr}
\end{figure}

The superconformal speed of sound of the NS matter ensures its stiffness needed to reach the masses of the PSR J0348+0432 \cite{Antoniadis:2013pzd} and PSR J0470+6620 \cite{dittmann:2024:10215108}. 
This is seen from Fig. \ref{fig:mr}, which shows the dependence of the NS mass on its radius and central density.
For the most probable scenarios the maximum mass is $2.0-2.2\rm M_\odot$.
These values are smaller than in the case of the two-flavor chiral quark model with local vector repulsion \cite{Ayriyan:2024zfw}.
This is due to the subdominant character of the nonlocal vector repulsion at high densities, which softens the quark EOS.
Nevertheless, the most probable scenarios agree with the high mass of the black widow pulsar PSR J0952-0607 \cite{Romani:2022jhd} corrected according to its rotation frequency (see footnote on page 9).
An important conclusion, which can be drawn from Fig. \ref{fig:mr}, is that the most probable onset masses of quark matter in NSs are below $1\rm M_\odot$.
This is a direct consequence of the small onset density of quark matter discussed above.
Phenomenologically, this means that all the observed NSs contain quark cores in their interiors.
It is remarkable that this scenario is statistically favorable compared to the purely hadronic NSs.
Our findings suggest that the most probable radius of a $1.4\rm M_\odot$ canonical NS is $\rm R_{1.4}=11.0-12.0~km$.
It is also seen from Fig. \ref{fig:mr} that the central density of the heaviest NSs is likely to not exceed $\varepsilon\simeq1250~\rm MeV/fm^3$, which is significantly smaller than $\varepsilon\simeq1430~\rm MeV/fm^3$ corresponding to the benchmark scenario of purely hadronic NSs.

\section{Conclusions}
\label{concl}

We have performed a physics-informed Bayesian analysis of the observational data on the masses, radii, and tidal deformability of NSs within the scenario when their interiors undergo a first-order phase transition to quark matter modeled within the framework of Maxwell construction of hybrid EOS.
The used scheme allows us to distinguish between the scenarios of NSs with quark cores and without them, which is its important advantage compared to the model-agnostic Bayesian analysis.
The results of the analysis are confronted to the baseline scenario of purely hadronic NSs and are shown to favor hybrid NSs with quark cores in their interiors.

To exclude systematic uncertainty due to unreliable modeling of hadronic matter, we utilized the microscopic DD2Y-T EOS, which agrees with the low-density constraints of $\chi$EFT, reproduces the properties of the ground state of normal nuclear matter and includes hyperonic degrees of freedom.
The three-flavor quark matter is modeled within a version of the NJL model with nonlocal interaction. 
In the deconfinement region the most important interaction channels are strong vector repulsion and diquark pairing controlled by the dimensionless couplings $\eta_V$ and $\eta_D$ of the microscopic Lagrangian.
Although vector repulsion ensures the stiffness of dense matter required to reach NS masses as high as at least $2\rm M_\odot$, the pairing of diquarks leads to the formation of a color-superconducting CFL phase.
The latter is important to provide not too late onset of deconfinement even in the case of very stiff quark matter, which is required for the TOV-stability of the quark branch of the NS mass-radius relation.
For simplicity, the analysis is limited to the case where the three quark flavors are degenerate in mass.
However, this does not introduce much uncertainty since in the deconfinement region bare quark masses do not play an important phenomenological role \cite{Ivanytskyi:2024zip}.

A new aspect of the study is the asymptotically conformal behavior of the quark matter EOS, which is provided by the nonlocal character of the interaction among quarks.
This element of the model automatically provides its consistency with the high-density constraints of perturbative QCD.
Moreover, regardless the values of the vector and diquark couplings, the conformal value of the squared speed of sound is reached from below, in agreement with the results of perturbative QCD. 

The vector and diquark couplings, which provide a proper Maxwell crossing of the quark and hadron EOSs, are found to belong to a wedge-shaped region in the $\eta_V-\eta_D$ plane.
This region is limited from right-below by the condition that the quark-hadron phase transition should not occur below the nuclear saturation density.
A narrow region in the vicinity of this low-density boundary represents the twin configurations of NSs.
The upper left boundary of the wedge-shaped region occurs because the phase transition is not possible. 
In a narrow region in the vicinity of this high-density boundary the phase transition either occurs at energy densities beyond the value in the center of the maximum mass configuration of the purely hadronic NSs, or at densities close to it where the quark branch of the corresponding mass-radius relation is TOV-unstable.

The Bayesian analysis of two sets of the observational data allowed us to find the posterior probability distribution of the vector and diquark couplings.
The basic set is given by the mass-radius data from the PSR~J0030+0451~\cite{miller:2019:3473466}, PSR~J0740+6620~\cite{dittmann:2024:10215108}, PSR~J0437–4715~\cite{choudhury:2024:13766753}, and the tidal deformability data from the GW170817~\cite{ligoLIGOP1800115v12GW170817}, while the full set additionally includes the mass-radius data from the PSR J0952-0607 \cite{Romani:2022jhd} and HESS~J1731-347~\cite{doroshenko:2023:8232233}.
Similar to the previous findings of Refs. \cite{Baym:2019iky,Gartlein:2023vif,Gartlein:2024cbj}, 
the found 60\% and 90\% credibility regions elongate in the $\eta_V-\eta_D$ plane along contours with positive slopes.
For the basic and full sets of the observational data, these regions concentrate around the most likely values of the couplings $\eta_V\simeq0.64$ and $\eta_D\simeq0.36$, and $\eta_V\simeq0.82$ and $\eta_D\simeq0.40$, respectively. 
Unlike the NJL model with local vector repulsion \cite{Ayriyan:2024zfw}, the 60\% and 90\% credibility regions do not include $\eta_V\lesssim 0.5$ for the basic and $\eta_V\lesssim 0.6$ for the full sets.
This is because in the nonlocal case considered in this work the vector repulsion is subdominant at high densities, which makes small vector couplings inconsistent with high NS masses.
The main modifications related to using the full set of observational data compared to the basic set correspond to increasing the most probable values of the vector and diquark couplings and narrowing the 60\% and 90\% credibility regions.

The most probable EOSs, which belong to the 60\% credibility region, have quark-hadron phase transitions at the energy density $150-300~\rm MeV/fm^3$. 
It is remarkable that the corresponding onset masses of quark matter in NSs are well below $1.0\rm M_\odot$.
This suggests that all the observed NSs have quark cores. 
The squared speed of sound of the most probable EOSs exhibits a peak of the amplitude $c_S^2\simeq 0.40$ located in the quark phase at $\varepsilon\simeq 500~\rm MeV/fm^3$.
The central densities of the heaviest NSs do not exceed $1250~\rm MeV/fm^3$, when $c_S^2\simeq 0.38$.
This signals that the behavior of quark matter, which can exist in the NS interiors, is firmly superconformal.
This allows NSs with CFL cores to reach maximum masses with the most probable values around $2.2\rm M_\odot$.

The most important finding of the work is that the posterior probabilities of the most likely hybrid quark-hadron EOSs significantly exceed the probability corresponding to the benchmark purely hadronic EOS.
The ratio of these probabilities is 18.72 for the basic and 276.15 for the full set of the observational data.
This provides statistical evidence that the scenario of NSs with CFL quark cores is favored compared to the scenario of purely hadronic NSs.

\vspace{0.5cm}
\section*{ACKNOWLEDGMENTS}
The authors acknowledge the support by NCN under grant No. 2021/43/P/ST2/03319.
The work of O.I. was performed within the program Excellence Initiative--Research University of the University of Wrocław of the Ministry of Education and Science. 

\vspace{0.5cm}
\section*{DATA AVAILABILITY}

The data supporting the findings of this article are openly available under the condition of citing the source \cite{Ayriyan_2025_17024839}.


\begin{appendix}
\section{Thermodynamic potential of three-flavor nonlocal NJL model}
\label{appA}

The Hubbard-Stratonovich transformation of the partition function $\mathcal{Z}$ is given by the identity
\begin{eqnarray}
\mathcal{Z}&=&\int\mathcal{D}\overline{q}\mathcal{D}q\exp\left(\int dx~\mathcal{L}\right)\nonumber\\
\label{A1}
&=&\int
\mathcal{D}\overline{q}\mathcal{D}q
\mathcal{D}\sigma_a
\mathcal{D}\omega_\mu\mathcal{D}d_{ab}^*\mathcal{D}d_{ab}^{ }
\exp\left(\int dx~\mathcal{L}^{\rm bos}\right),
\quad\quad
\end{eqnarray}
The exponential in this expression includes the intergation over the four-space of the volume $\beta V$, where $\beta=1/T$ is the inverse temperature and $V$ stands for the three-volume. 
The auxiliary bosonic fields $\sigma_a$, $\omega_\mu$ and $d_{ab}$ are conjugated to the quark bilinears in the scalar, vector and diquark interaction channels, and enter the bosonized Lagrangian as
\begin{eqnarray}
\mathcal{L}^{\rm bos}\hspace*{-.1cm}&=&\overline{q}(i\slashed\partial-m+\mu\gamma_0)q
\nonumber\\
&-&\sum_{a=\overline{0,8}}\left(s_a\sigma_a+\frac{\sigma_a\sigma_a}{4G_S}\right)
+j_\mu\omega^\mu+\frac{\omega_\mu\omega^\mu}{4G_V}\nonumber\\
\label{A2}
&-&\sum_{a,b=2,5,7}\hspace*{-.2cm}
\left(\frac{d_{ab}^+\Delta_{ab}^{ }+\Delta_{ab}^*d_{ab}^{ }}{2}+
\frac{\Delta_{ab}^*\Delta_{ab}^{ }}{12G_D}\right).
\end{eqnarray}
It gives a direct access to the Euler-Lagrange equations for the bosonic fields.
Within the mean field approximation applied in Ref. \cite{Ivanytskyi:2024zip}, these fields are replaced by the expectation values obtained from the Euler-Lagrange equations.
The corresponding functional integrations are supressed.
Equivalently, these expectation values can be obtained by minimizing the thermodynamic potential.

Noticing that charge conservation requires the commutation of the mean-field Lagrangian with the charge matrix $\hat Q = {\rm diag}(2/3,-1/3,-1/3)$ \cite{Hell:2009by}, we conclude the absence of the non-diagonal flavor group generators in the mean-field Lagrangian, i.e. vanishing of the expectation values of the corresponding scalar currents and scalar fields.
This is also the case for the scalar fields connected to the traceless flavor generators $\tau_3$ and $\tau_8$.
Thus, $\langle\sigma_a\rangle=\delta_{a0}\sigma$ under the mean field approximation. 
Applying the proper Lorentz transformation, we bring the expectation value of the vector field to the form $\omega_\mu=g_{\mu0}\omega$. 
Only three diquark fields survive the averaging in the ground state due to the locking of the color and flavor indexes of quarks in the CFL matter \cite{Pisarski:1998nh}.
We utilize a color-flavor rotation, at which the surviving fields have a diagonal structure $a=b$ \cite{Buballa:2003qv}.
Furthermore, given the flavor symmetry of the CFLL phase, we conclude that the expectation values of these diquark fields coincide.
Based on this we introduce
$\Delta\equiv\langle\Delta_{11}\rangle+\langle\Delta_{22}\rangle+\langle\Delta_{33}\rangle$ so that $\langle\Delta_{ab}\rangle=\delta_{ab}\Delta/3$.
In general, $\Delta$ is complex. 
But it enters the mean-field Lagrangian via the product with its complex conjugate $\Delta^*$.
As a result, only the modulus of this quantity appears in the expression for the thermodynamic potential.
Therefore, below $\Delta$ stands for this modulus, while its sign is suppressed for shortening the notations.
With this the mean-field Lagrangian becomes
\begin{eqnarray}
\label{A3}
\mathcal{L}_{\rm MF}+\mu~q^+q=
\overline{\mathcal{Q}}\hat{\mathcal{S}}\mathcal{Q}-
\frac{\sigma^2}{4G_S}+\frac{\omega^2}{4G_V}-\frac{\Delta^2}{4G_D}.
\end{eqnarray}
Here $\mathcal{Q}^T=(q,q^c)/\sqrt{2}$ and $\hat{\mathcal{S}}$ are the Nambu-Gorkov bispinor and propagator, respectively, while $q^+q$ is operator of the conserved quark number density conjugated to the quark chemical potential $\mu$. 
In the momentum representation the inverse Nambu-Gorkov propagator reads
\begin{eqnarray}
\label{eq:NGpropagator}
\hat{\mathcal{S}}^{-1}=\left(
\begin{array}{ll}
\hspace*{.6cm}S^{-1}_+\hspace*{.7cm}i\Delta^{ }g_k\gamma_5\mathcal{O}\\
i\Delta^*g_k\gamma_5\mathcal{O}\hspace*{.9cm}S^{-1}_-
\end{array}\right),
\end{eqnarray}
where $S^{-1}_\pm=\slashed k-M_{\bf k}\pm\gamma_0(\mu+\omega g_{\bf k})$, the zeroth component of the four-momentum $k$ represents the Matsubara frequencies of fermions, i.e. $k_0=i(2n+1)\pi T$, $M_{\bf k}=m+\sigma g_{\bf k}$ is the effective momentum dependent quark mass introduced in Sec. \ref{sec2}, and the operator $\mathcal{O}=\tau_2\lambda_2+\tau_5\lambda_5+\tau_7\lambda_7$ acts in the color-flavor space and is introduced for shortening the notations.

The single quark energies can be found by solving the equation ${\rm det}~\hat{\mathcal{S}}=0$ with respect to $k_0$.
This is equivalent to analyzing the nonvanishing elements of the diagonalized Nambu-Gorkov propagator.
Such a diagonalization is provided by a rotation in the color-flavor space.
The corresponding quark basis is given not by their color-flavor states but by an octet of degenerated states and a singlet state \cite{Buballa:2003qv}.
The single particle energies and degeneracies of these octet and singlet states, which are introduced in Sec. \ref{sec2}, are labeled by the subscript index ``$j$''.

The mean-field Lagrangian (\ref{A3}) is quadratic in quark fields, which allows integrating them out.
The resulting statistical partition is
\begin{eqnarray}
    \ln\mathcal{Z}&=& \ln\int
    \mathcal{D}\overline{\mathcal{Q}}~
    \mathcal{D}\mathcal{Q}
    \exp\left(\int dx~(\mathcal{L}_{\rm MF}+\mu \overline{q}q)\right)\nonumber\\
    &=&
    \frac{1}{2}{\rm tr}\ln\hat{\mathcal{S}}-
    \beta V\left(
    \frac{\sigma^2}{4G_S}-\frac{\omega^2}{4G_V}+\frac{\Delta^2}{4G_D}\right),
\end{eqnarray}
where the factor $1/2$ in the first term compensates the artificial Nambu-Gorkov doubling of degrees of freedom.
Performing the trace over the Nambu-Gorkov, Dirac, color, flavor, three-momentum, and Matsubara indexes, and using the thermodynamic identity $\Omega=-\ln\mathcal{Z}/\beta V$, we arrive at the thermodynamic potential
\begin{eqnarray}
    \Omega&=&-\sum_{j,a}d_j\int\frac{d\bf k}{(2\pi)^3}\left[\frac{\epsilon_{j\bf k}^a}{2}-
    T\ln(1-f{_{j\bf k}^a})\right]
    \nonumber\\
    \label{A6}
    &+&
    \frac{\sigma^2}{4G_S}-\frac{\omega^2}{4G_V}+\frac{\Delta^2}{4G_D}.
\end{eqnarray}
where $f_{j\bf k}^a=\left(1+e^{-\epsilon_{j\bf k}^a}\right)^{-1}$ is the single particle distribution function of quarks ($a=+$) and antiquarks ($a=-$).
The final step in deriving Eq. (\ref{eq:Omega}) corresponds to taking the zero-temperature limit.
For this, we notice that 
\begin{eqnarray}
\label{A7}    
T\ln\left(1-f_{j\bf k}^a\right)|_{T\rightarrow0}=
\epsilon_{j\bf k}^a\theta\left(-\epsilon_{j\bf k}^a\right).
\end{eqnarray}
Inserting this identity into Eq. (\ref{A6}) we arrive at the zero temperature thermodynamic potential from Sec. \ref{sec2}.

\section{Competing constraints at $2\rm M_\odot$}
\label{appB}

\begin{figure}[t]
    \includegraphics[width=\columnwidth]{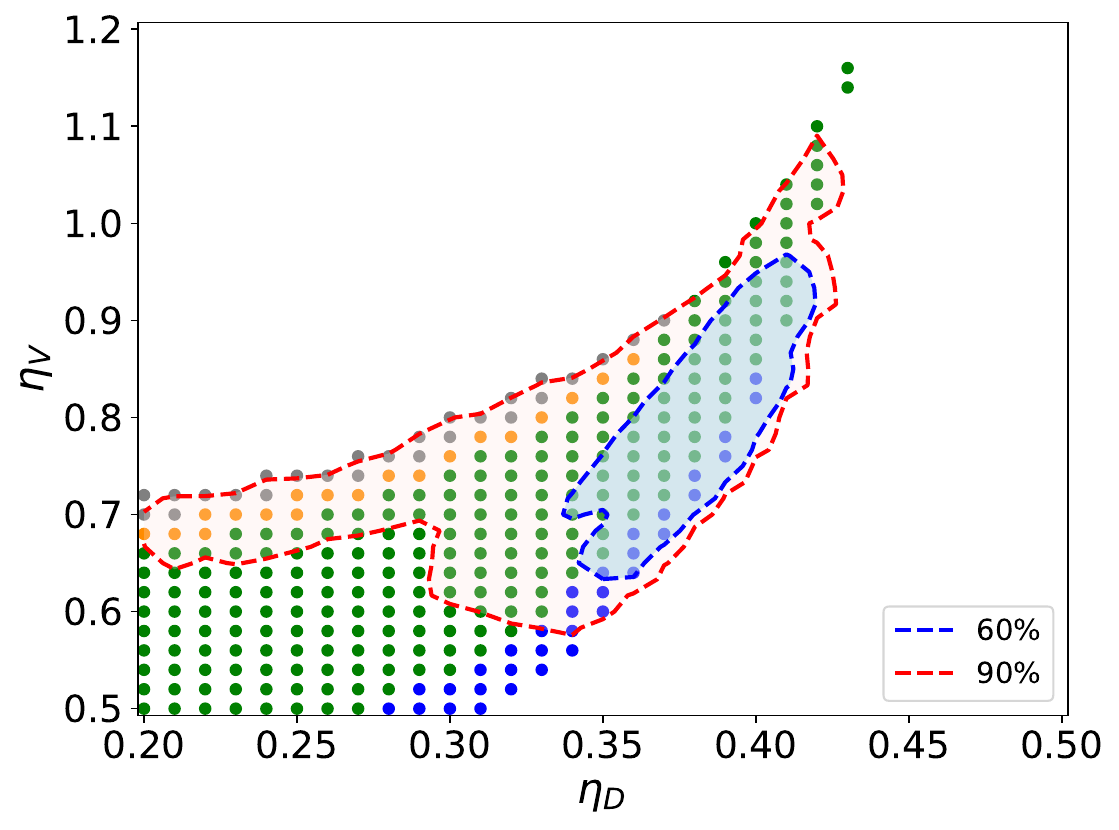}
    \caption{The same as in Fig. \ref{fig:BA2} but with contours of the 60\% and 90\% confidence intervals obtained from the Bayesian analysis of the full set of the observational data supplemented with the constraint from the precise mass measurement on the $2\, \rm M_\odot$ pulsar J0348+0432 \cite{Antoniadis:2013pzd} which comes without a constraint for its radius.
    }
    \label{fig:appB}
\end{figure}

In this appendix, we want to illustrate the effect that using the precise mass measurement of the $2\rm M_\odot$ pulsar J0348+0432 \cite{Antoniadis:2013pzd}, which comes without a constraint for its radius, along with the NICER combined mass-radius measurement for the other $2\rm M_\odot$ pulsar J0740 + 6620, has on the $60\%$ and $90\%$ confidence regions in the $\eta_V$ - $\eta_D$ parameter plane.
We show the result in Fig. \ref{fig:appB}.

Comparison of this result with both panels of Fig.~\ref{fig:BA2} in Sect. \ref{sec6} 
reveals the counter-intuitive effect that adding another mass constraint does not narrow the posterior confidence regions, as one could naively expect, but actually widens it! 
This concerns in particular the $90\%$ confidence region (shown by a red dashed line), which now includes all gray and yellow marked parameter regions that would correspond to purely hadronic NS sequences in the $M-R$ diagram. In Fig.~\ref{fig:BA2}, these purely hadronic sequences are0 not included in the $90\%$ confidence region, and thus are clearly disfavored against the hybrid star sequences.
How to explain this widening of the confidence region?
Adding a $2\,M_\odot$ pulsar mass that is not constrained with respect to the radius, the hadronic baseline sequences at high mass, which did not satisfy the NICER constraint on the minimal radius $R_{2.0} < 11.79$ km at that mass, gain an additional probability from the less restrictive mass-only constraint of the PSR J0348 + 0432. 
We, therefore, suggest that once a more restrictive combined mass-radius measurement like the ones provided by NICER exists, one should not neutralize the selective power of this measurement by an unconstrained mass-only measurement, even if it is rather precise.

\end{appendix}

\bibliography{bibliography}

\onecolumngrid

\twocolumngrid

\end{document}